\documentclass{elsart}
\usepackage{amsmath,color,amsfonts,amssymb,afterpage,hyperref}
\usepackage{graphicx}
\usepackage{multirow}

\graphicspath{{eps/}{inkscape/}}

\def\text#1{{\rm #1}}

\def\Re{\mathbb R}

\def\Z{\mathbb Z}

\def\pd{\partial}
\def\ji{\varphi}
\def\Mfold{\mathcal M}

\def\eqref#1{(\ref{#1})}
\def\FETK{{\sc FETK}}

\numberwithin{equation}{section}

\begin{document}
%=============================================================================
\begin{frontmatter}

\title
{Solving the Einstein constraint equations \\
on multi-block triangulations using \\
finite element methods}

\author[lsu-ph,lsu-cct]{Oleg Korobkin}
\author[lsu-mt,lsu-cct]{Burak Aksoylu}
\author[ucsd]{Michael Holst}
\author[lsu-ph,lsu-cct]{Enrique Pazos}
\author[umd-ph,umd-cscmm]{Manuel Tiglio}
\address[lsu-ph]{Department of Physics and Astronomy, Louisiana State University}
\address[lsu-mt]{Department of Mathematics, Louisiana State University}
\address[lsu-cct]{Center for Computation and Technology, Louisiana State University}
\address[ucsd]{Department of Mathematics, University of California at San Diego.}
\address[umd-ph]{Department of Physics, University of Maryland}
\address[umd-cscmm]{Center for Scientific Computation and Mathematical Modeling, 
         University of Maryland.}

\date{\today}

%-----------------------------------------------------------------------------
\begin{abstract}
  In order to generate initial data for nonlinear relativistic simulations,
  one needs to solve the Einstein constraints, which can be cast into a coupled
  set of nonlinear elliptic equations. Here we present an approach for solving
  these equations on three-dimensional multi-block domains using finite
  element methods.
  We illustrate our approach on a simple example of Brill wave initial data,
  with the constraints reducing to a single linear elliptic equation for the
  conformal factor $\psi$.
  We use quadratic Lagrange elements on semi-structured simplicial
  meshes, obtained by triangulation of multi-block grids. In the case of
  uniform refinement the scheme is superconvergent at most mesh vertices, due
  to local symmetry of the finite element basis with respect to local spatial
  inversions. We show that in the superconvergent case subsequent unstructured
  mesh refinements do not improve the quality of our initial data.
  As proof of concept that this approach is feasible for generating multi-block
  initial data in three dimensions, after constructing the initial data we
  evolve them in time using a high order finite-differencing multi-block
  approach and extract the gravitational waves from the numerical solution.
\end{abstract}
%-----------------------------------------------------------------------------

\end{frontmatter}

%=============================================================================
\section{Introduction}
\label{S:Intro}
%-----------------------------------------------------------------------------

This paper is part of an effort to numerically solve Einstein's equations on 
general domains with arbitrary shapes and topologies. Accurate simulations on
such domains require multiple numerical grids covering different areas of the
domain, possibly refined where the solution has interesting features. For the
numerical discretization one can use either finite differences (FD), spectral
methods, or finite elements (FE). Current research in numerical relativity
community is dominated by finite differences methods, not the last
reason for that being the relative ease of finite differences implementation
and parallelization.  A lot of important results in areas of hyperbolic
solvers, boundary conditions, domain representation, apparent horizon finders,
wave extraction, stability proofs etc., were established and successfully
implemented using the FD methods
(see~\cite{Lehner01a,Pretorius:2005gq} and references therein).

Our approach was designed to allow for fast
and accurate transformation of the numerical function between FE and FD
representations. We demonstrate that the FE solution to an elliptic equation,
generated for the given FD grid, is sufficiently accurate and convergent for
3D relativistic simulations with independent finite difference hyperbolic
solver (QUILT).

Spectral methods are also widely
used~\cite{Bonazzola-etal-1996:spectral-methods-in-gr,lrr-2009-1}, mainly for 
solving elliptic equations, such as those arising in initial value problems
\cite{Pfeiffer:2002wt,Cook00a,Pfeiffer:2000um,Kidder99a,Pfeiffer:thesis}, 
constrained evolution~\cite{Bonazzola98a} and apparent horizon
finders~\cite{Lin:2007cd,Nakamura84}.
They can as well be applied to hyperbolic problems~\cite{Tadmor94}, as in the
recent very successful simulations of the binary black holes
merger~\cite{Scheel:2008rj} (using an approach, developed by the
Cornell-Caltech
collaboration~\cite{Kidder00a,Kidder00b,Scheel2002a,Scheel-etal-2006:dual-frame}).
Spectral methods can produce exceptionally accurate results, but in order to
use these results in a FD evolution code one needs to perform an additional
interpolation step. The interpolation can be too slow for some applications,
for instance, in constrained evolution schemes. In our approach, the
interpolation is trivial, and the number of degrees of freedom on FE and FD
grid is the same, which allows to exchange solution between the FE and FD
solvers without reducing its accuracy by interpolation. Another advantage of
FE methods compared to spectral methods is that the matrices resulting from FE
discretizations are sparse.

Finite element methods can also be applied to hyperbolic problems,
however, stable and accurate discretizations require using discontinuous
Galerkin (DG) flavor of finite elements (otherwise the approximation
is suboptimal: order of convergence falls to $(p-1)$ and the solution is not
necessarily stable). 

% /---------------------------------------------------------------------\
  \begin{figure}[!htp]
  \begin{center}
  \includegraphics[height=0.2\textheight]{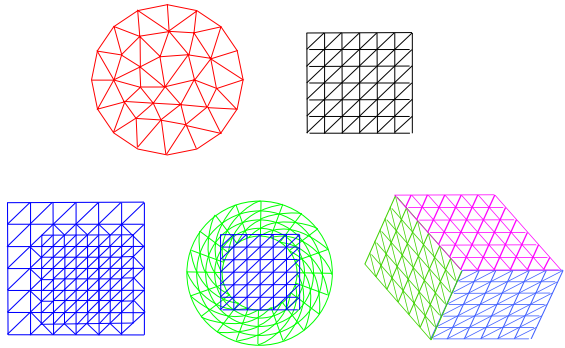}
  \caption{
    Types of FE meshes.
    Top left: unstructured mesh,
    top right: regular mesh,
    bottom: semistructured meshes.
  }
  \label{F:semistr}
  \end{center}
  \end{figure}
% \---------------------------------------------------------------------/

In this paper, the term \emph{grid} is used in reference to the finite
differences grid (an ordered set of points), as opposed to the term
\emph{mesh} for the finite element triangulation. Domain triangulations can be
classified as \emph{regular}, \emph{unstructured} or \emph{semistructured},
depending on their degree of symmetry (see figure~\ref{F:semistr}). Finite element
methods can handle domains of arbitrarily complex shapes and topologies, since
every geometrical shape admits approximation with succesively refined
unstructured symplectic triangulations. However, the domains encountered in
numerical relativity are relatively simple in the sense that they can be
covered with semistructured meshes. We argue in favor of using semistructured
meshes over unstructured ones. One of the motivations is that they can be
quickly constructed on top of existing FD grids without the need to explicitly
maintain all the information about mesh elements, such as edges, faces etc.
Raising the order of finite elements, one can essentially achieve the spectral
convergence; however, here we concentrate on lower-order finite elements
(first- and second-order), and show how the \emph{superconvergence} on
semistructured grids for second-order elements can be exploited to obtain
accurate 4-th order convergent solution, suitable for use in finite
differences simulation. 

The method is tested on the particular example of \emph{multipatch} FD grid.
In differential geometry it is common to define a manifold through a set of
possibly overlapping patches, with each patch mapped into an open, 
simply-connected subset of Euclidean space (see, for example,~\cite{Wald84}).
This is a natural way of describing manifolds with nontrivial topology, which
cannot be covered by a single coordinate chart. After discretization every
chart becomes a FD grid, which can be thought of as a discrete coordinate
chart, mapped into a region of $\Z^3$.
At the continuum level, the patches are glued together by coordinate
transformations in the areas where they intersect. At the discrete level, 
if the neighboring patches do not have common points interpolation can be used 
to fill in the missing ones; this approach is commonly referred to as
\emph{overlapping-grids}. On the other hand, in a \emph{multi-block} approach
the patches abut rather than overlap, and the grids are constructed in such a
way that neighboring grids share boundary points.

A multipatch approach in numerical relativity has several advantages. In many
cases the domain of interest is asymptotically flat. If it contains one or
more black holes, then each hole can be excised from the computational domain
by introducing an inner smooth boundary around the singularity. If
appropriately chosen, this boundary does not require any physical boundary
conditions, since all characteristic modes leave the domain. It is
preferable that this boundary is smooth, which in general requires the use of
multiple patches. Similarly, the preferred shape for the outer boundary when
simulating asymptotically flat spacetimes is a sphere, as this is the topology
of null infinity $\mathcal{J}^{+}$, which is best suited for extracting
gravitational waves. A multipatch domain structure easily accomodates both
type of boundaries while avoiding coordinate singularities, such as those
associated with spherical or cylindrical coordinates. The use of multiple
patches is unavoidable in cosmological simulations with non-trivial
topologies. In addition, multiple patches are in general more efficient than
regular grids, since they decouple different spatial directions. For example,
under conditions of approximate spherical symmetry, one could surround the
system with spherical patches and fix the number of points in angular
direction, while increasing the domain only in radial direction. This way,
pushing the outer boundary out becomes an order $N$ problem, as opposed to
$N^3$ with Cartesian grids.  This same feature makes them useful, in
particular, for many relativistic astrophysical studies which are assumed to
be approximately spherically symmetric \cite{Zink07:MPAstro}.

Einstein's equations are often written in a form such that the equations
divide into hyperbolic (evolution) and elliptic (constraint) sub-systems. When
solving the Einstein {\em vacuum} (that is, in the absence of matter)
evolution equations, these can be cast in linearly-degenerate form, and the
solutions are in general expected to be smooth. Those cases are ideally suited
for high order or spectral methods.  When including matter, on the other hand,
for example when dealing with the general relativistic hydrodynamics
equations, shocks are expected. In those cases a possible choice is  high
resolution shock capturing methods~\cite{lrr-2003-4,Marti99}  adapted to the
presence of several patches for the hydrodynamical sub-system and high order
or spectral methods for the metric
sub-system~\cite{Zink07:MPAstro,Duez07:RFluid}.

This paper is concerned with the elliptic sector of Einstein's equations, more
precisely with the generation of {\em initial data} needed for evolutions on 
multipatch geometries. 
Initial data on a spacelike hyperslice has to satisfy a set of
Enistein's constraints, which can be cast into a coupled nonlinear elliptic
system of partial differential equations (PDE).
This elliptic PDE system has been studied extensively for many years,
with a complete solution theory developed in the case of domains which
are closed (i.e. compact without boundary) 3-manifold spatial slices of spacetime 
with constant or nearly-constant mean extrinsic curvature~\cite{jI95,jIvM96}.
More recently, this solution theory has been extended to both closed
3-manifolds and to compact 3-manifolds with boundary,
having mean extrinsic curvature
far from constant~\cite{HKN07a,HNT07b,HNT07a}.
It has also been shown recently~\cite{Hols2001a,HoTs07a,HoTs07b} that
geometric partial differential equations of this type can be solved 
accurately and efficiently using adaptive finite element methods
on manifolds with general topologies. For general multi-block systems, the
additional structure allows one to construct semi-structured triangulations, 
which generate superconvergent finite element solutions. The property of
superconvergence holds at the vertices of the multi-block grid triangulation,
which simplifies transport of the solution to the finite difference grid, since
no interpolation is required. 

The above considerations define the approach of
this paper, which involves solving the Einstein constraint equations on 
semi-structured multi-block triangulations using finite element methods.
We apply this approach to the Brill wave initial data
problem, which represents a linear vacuum case with zero extrinsic
curvature, with the constraints system reducing to a single elliptic equation
$\Delta\psi-V\psi=0$ for the conformal factor $\psi$ (where $\Delta$ is the
flat 3-dimensional laplacian and V is a potential function related to the Ricci
scalar of the freely given metric). Nevertheless, the results presented are
expected to hold in general case, because the property of superconvergence is
the property of the mesh and the finite element spaces used. If this property
holds for a linearized problem, it is also expected to hold for the full
nonlinear one (see Chapter 9 of~\cite{Wahlbin95:superconvergence}).

The paper is organized as follows. Section \ref{S:SolvFEtk} gives an overview
and summary of the Finite Element Toolkit (FETK)~\cite{Hols2001a}, 
which we here use for solving the Einstein constraint equations with
finite element methods.
Section \ref{S:GenMesh} discusses our approach to multi-block
triangulations and, in particular, why superconvergence is expected. In
section \ref{S:QualSol} we evaluate the accuracy of our approach by solving
for several three-dimensional elliptic test problems on multi-block domains
with known exact solutions. Finally, in section \ref{S:BrillWaveID} we solve
the Einstein constraints on a multi-block domain for the case of Brill
waves. After constructing initial data, we use the multi-patch infrastructure
QUILT to solve for the Einstein evolution equations in time and extract
gravitational waves from the numerical solution. QUILT is an ongoing effort,
more details of which can be found in
\cite{Schnetter:2006pg},
\cite{lehner-2005-22},
\cite{diener-2007-32},
\cite{Dorband05a},
\cite{Pazos:2006kz},
\cite{Zink07:MPAstro}.
In tensor notations, used throughout the paper, Latin indices from the
beginning of the alphabet ($a$,$b$,...) run from 0 to 3, Latin indices $i$,
$j$, etc. run from 1 to 3, and the usual Einstein summation rule is assumed.

%=============================================================================
\section{Solving elliptic PDEs using the Finite Element ToolKit}
\label{S:SolvFEtk}
%-----------------------------------------------------------------------------

In this paper we use the Finite Element 
ToolKit (FETK)~\cite{Hols2001a} 
(see also~\cite{BaHo02a,HoTs07b}) to solve the Einstein
constraint equations. 
\FETK{}  is an adaptive multilevel finite element code developed 
over a number of years by the Holst research group at UC San Diego 
and their collaborators (see also~\cite{website:fetk-org}).
It is designed to produce provably accurate numerical solutions to
nonlinear elliptic systems of tensor equations on (Riemannian)
multi-dimensional manifolds in an optimal or nearly-optimal way.
We will summarize the main features of~\FETK{} here; more detailed
discussions of its use for general geometric PDE may be
found in~\cite{Hols2001a,HoTs07a}, and specific application to the
Einstein constraint equations may be found in~\cite{HoTs07b,ABBH07a}.

\FETK{} contains an implementation of a ``solve-estimate-refine'' algorithm,
employing inexact Newton iterations to treat non-linearities.   
The linear Newton equations at each
inexact Newton iteration are solved with unstructured algebraic
multilevel methods which have been constructured to have
optimal or near-optimal space and time complexity (see~\cite{AkHo02,ABH02a}).
The algorithm is supplemented with a
continuation technique when necessary.  
\FETK{} employs {\em a posteriori} error
estimation and adaptive simplex subdivision to produce
provably convergent adaptive solutions (see~\cite{CHX06a,CHX06b}).

Several of the features of \FETK{} are somewhat unusual, some of which are:

\begin{itemize}
\item {\em Abstraction of the elliptic system}: The elliptic system 
     is defined only through a nonlinear weak form along with an associated
     linearization form over the domain manifold. 
     To use the {\em a posteriori} error estimator, a third function
     $F(u)$ must also be provided (essentially the strong form of the problem).
\item {\em Abstraction of the domain manifold}: The domain manifold is
     specified by giving a polyhedral representation of the topology, along
     with an abstract set of coordinate labels of the user's interpretation,
     possibly consisting of multiple charts.
     \FETK{} works only with the topology of the domain, the connectivity
     of the polyhedral representation.
\item {\em Dimension independence}: The same code paths are taken for
     two-,  three- and higher-dimensional problems. 
     To achieve this dimension independence, \FETK{} employs the simplex as its 
     fundamental geometrical object for defining finite element bases.
\end{itemize}

%-----------------------------------------------------------------------------
\subsection{Weak Formulation Example}
\label{SS:WeakForm}
%-----------------------------------------------------------------------------

We give a simple example to illustrate how to construct 
a weak formulation of a given PDE. 
Here we assume the 3--metric to be flat so that $\nabla$ is the ordinary
gradient operator and $\cdot$ the usual inner product.  
Let $\Mfold$ represent a connected domain in $\Re^3$ with a smooth
orientable boundary $\pd\Mfold$, formed from two disjoint 2-dimensional
surfaces $\pd_0\Mfold$ and $\pd_1\Mfold$.

Consider now the following semilinear equation on $\Mfold$: 
\begin{eqnarray}
  \label{eqn:strong}
- \nabla \cdot (a(x) \nabla u(x)) + b(x,u(x)) 
      & = & 0 ~\text{~in~} \Mfold, \\
  \label{eqn:strong_robin}
n(x) \cdot (a(x) \nabla u(x)) + c(x,u(x))
      & = & 0 ~\text{~on~} \partial_1 \Mfold, \\
  \label{eqn:strong_dirichlet}
 u(x) & = & f(x) ~\text{~on~} \partial_0 \Mfold,
\end{eqnarray}
where $n(x) : \partial \Mfold \mapsto \Re^3$ 
is the unit normal to $\partial \Mfold$, and where
\begin{eqnarray}
a : \Mfold \mapsto \Re^{3 \times 3},
& \ \ \ &
b : \Mfold \times \Re \mapsto \Re, \\
c : \partial_1 \Mfold \times \Re \mapsto \Re,
& \ \ \ &
f : \partial_0 \Mfold \mapsto \Re.
\end{eqnarray}

If the boundary function $f$ is regular enough so that
$f \in H^{1/2}(\partial_0 \Mfold)$, then from the
Trace Theorem~\cite{Adam78}, 
there exists $\bar{u} \in H^1(\Mfold)$
such that $f = \bar{u}|_{\partial_0 \Mfold}$ in the sense of the
Trace operator (where $H^1(\Mfold)$ is the Hilbert space of all real
$L_2$-integrable functions on $\Mfold$ with $L_2$-integrable weak
derivative~\cite{R:Gok06}.)
Employing such a function $\bar{u} \in H^1(\Mfold)$,
the weak formulation has the form:
\begin{equation}
   \label{eqn:weak_model}
\text{~Find~} u \in \bar{u} + H_{0,D}^1(\Mfold)
   ~\text{~s.t.~}
     \langle F(u),v \rangle = 0,
~~\forall~ v \in H^1_{0,D}(\Mfold),
\end{equation}
where the nonlinear form is defined as:
\begin{equation}
\langle F(u),v \rangle = \int_{\Mfold} \left(a \nabla u \cdot \nabla v
    + b (x,u) v \right) ~dx
  + \int_{\partial_1 \Mfold} c(x,u) v~ds.
\label{Eq:NonlinearForm}
\end{equation}
The ``weak'' formulation of the problem given by
equation~(\ref{eqn:weak_model}) imposes only one order of
differentiability on the solution $u$, and only in the weak or
distributional sense.
Under suitable growth conditions on the nonlinearities $b$ and $c$,
it can be shown that this weak formulation makes sense, in that the
form $\langle F(\cdot),\cdot \rangle$ is finite for all arguments,
and further that there exists a (potentially unique) solution
to~(\ref{eqn:weak_model}).
In the specific case of the individual and coupled Einstein constraint
equations, such weak formulations are derived and analyzed
in~\cite{HKN07a,HNT07b}.

To analyze linearization stability, or to apply a numerical
algorithm such as Newton's method, we will need the
bilinear linearization form $\langle DF(u)w,v \rangle$,
produced as the formal Gateaux derivative of the nonlinear form
$\langle F(u),v \rangle$:
$$
\langle DF(u)w,v \rangle =
   \left.\frac{d}{d\epsilon}
 \langle F(u+\epsilon w),v \rangle \right|_{\epsilon=0}
$$
\begin{equation}
   \label{eqn:linearize_example}
= \int_{\Mfold} \left(a \nabla w \cdot \nabla v
+ \frac{\partial b(x,u)}{\partial u} w v \right) ~dx 
+ \int_{\partial_1 \Mfold}
  \frac{\partial c(x,u)}{\partial u} w v ~ds.
\end{equation}
Now that the nonlinear weak form $\langle F(u),v \rangle$
and the associated bilinear
linearization form $\langle DF(u)w,v \rangle$
are defined as integrals, they can be evaluated using
numerical quadrature to assemble a Galerkin-type discretization
involving expansion of $u$ in a finite-dimensional basis.

As was the case for the nonlinear residual
$\langle F(\cdot),\cdot \rangle$, the matrix
representing the bilinear form in the Newton iteration is easily assembled,
regardless of the complexity of the bilinear form 
$\langle DF(\cdot)\cdot,\cdot \rangle$.
In particular, the algebraic system for $w = \sum_{j=1}^n \beta_j \phi_j$
has the form:
\begin{equation} \label{algebraicSystem}
A U = F,
\ \ \ \ \ 
U_i = \beta_i,
\end{equation}
where
\begin{eqnarray}
A_{ij} &=& \langle DF( \bar{u}_h + \sum_{k=1}^n \alpha_k \phi_k )\phi_j,
         \psi_i \rangle, \\
F_i &=& \langle F( \bar{u}_h + \sum_{j=1}^n \alpha_j \phi_j),\psi_i \rangle.
\end{eqnarray}
and $\{\phi_i\}_{i=1}^N$, $\{\ji_j\}_{j=1}^N$ are the bases of the test
and trial $N$-dimensional finite element spaces.
As long as the integral forms
$\langle F(\cdot),\cdot \rangle$ and $\langle DF(\cdot)\cdot,\cdot \rangle$
can be evaluated at individual points in the domain, then quadrature can
be used to build the Newton equations, regardless of the complexity of
the forms.
This is one of the most powerful features of the finite element method,
and is exploited by \FETK{} to make possible the representation and
discretization of very general geometric PDEs on manifolds.
It should be noted that there is a subtle difference between the approach
outlined here (typical for a nonlinear finite element approximation) and
that usually taken when applying a Newton-iteration to a nonlinear finite
difference approximation.
In particular, in the finite difference setting the discrete equations
are linearized explicitly by computing the Jacobian of the system of 
nonlinear algebraic equations.
In the finite element setting, the commutativity of linearization and
discretization is exploited; the Newton iteration is actually
performed in function space, with discretization occurring
``at the last moment.''

%=============================================================================
\section{Generating semi-structured multi-block triangulations with
  superconvergent properties}
\label{S:GenMesh}
%-----------------------------------------------------------------------------

In order to avoid the extra step of interpolating the finite element numerical
solution to a multi-block grid, we use finite element meshes with
vertices located at the multi-block grid points. We construct such meshes by 
dividing a convex hull of the set of grid points into simplices with vertices
only at those points. This procedure of building a simplicial mesh based on 
a set of points is usually referred to as \emph{triangulation}. Delaunay's 
triangulation (see, for instance,~\cite{R:GeB98}) is an example of such a
procedure, generating meshes with simplices of the highest possible quality in
flat space. Although Delaunay's triangulation minimizes a condition number in a
resulting linear system, it is mostly used for sets of points of general
position since it generates completely unstructured meshes; for more regular grids
its algorithm is unnecessarily complex. Here we use a simpler and
more straightforward algorithm to generate semi-structured meshes with the
additional advantage of having superconvergent properties.

\begin{figure}[!htp]
  \begin{center}
  \includegraphics[height=0.2\textheight]{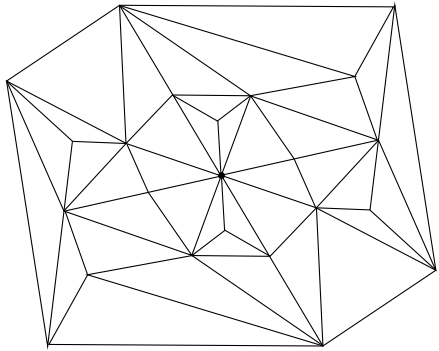}
  \caption{
    Example of two-dimensional simplectic mesh, locally symmetric with
    respect to the point ~\cite{R:SSW96}. }
  \label{F:supersym}
  \end{center}
  \end{figure}

The term {\it superconvergence} applies in various contexts, when the local or global
convergence order of a numerical solution is higher than one would 
expect~\cite{Chen82a:fem,ZhuLin98:hyperconvergence,Wahlbin95:superconvergence,KvrivzekNeittaanmaki87:global}.
The type of superconvergence we are interested in is superconvergence by local
symmetry~\cite{Wahlbin95:superconvergence}, which occurs in function values for
discretizations of second order elliptic boundary value problems, and amounts
to an additional $0 < \sigma\le1$ in the convergence order. 
It occurs at any given point in which the finite element basis 
is locally symmetric, or approximately symmetric, with respect to local spatial
inversion at that point (see figure~\ref{F:supersym}). The triangulation
method that we use produces simplicial meshes with this type of symmetry at
all vertices inside each block, and at some vertices at the interblock
boundaries. Therefore, in our meshes we expect superconvergence everywhere but
at the non-symmetric interblock boundary points.
For second order elliptic equations and piecewise polynomial finite elements
of even degree superconvergence occurs in the solution itself, while for odd
degree it occurs in the first derivatives of the
solution~\cite{R:SSW96,R:Sch98}.

% /---------------------------------------------------------------------\
% \---------------------------------------------------------------------/

With this in view, the quadratic Lagrange finite elements at the nodal
points of the meshes that we use are expected to give solutions with
convergence order $3+\sigma$ (for some $0 < \sigma\le1$).  This is an important
advantage of multi-block triangulations: the resulting grid solutions have
a convergence rate which is even higher than the global convergence rate of the
original finite element solution, almost everywhere.

We start by dividing a cell, then combine cells into a block, and finally
combine blocks together in a conforming way. Next we explain how these
procedures are performed.
The main complexity in building multi-block triangulations comes from the last
step, since triangulations of each block are not necessarily conforming at
their interfaces. We will show, however, that our method guarantees conforming
simplicial triangulations.

\begin{figure}[!htp]
  \begin{center}
  \includegraphics[height=0.1\textheight]{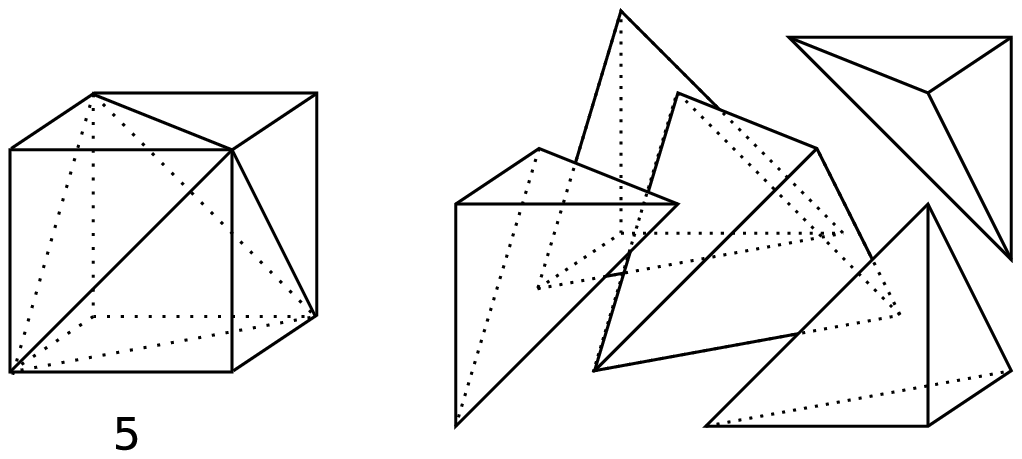}
  \\
  \includegraphics[height=0.1\textheight]{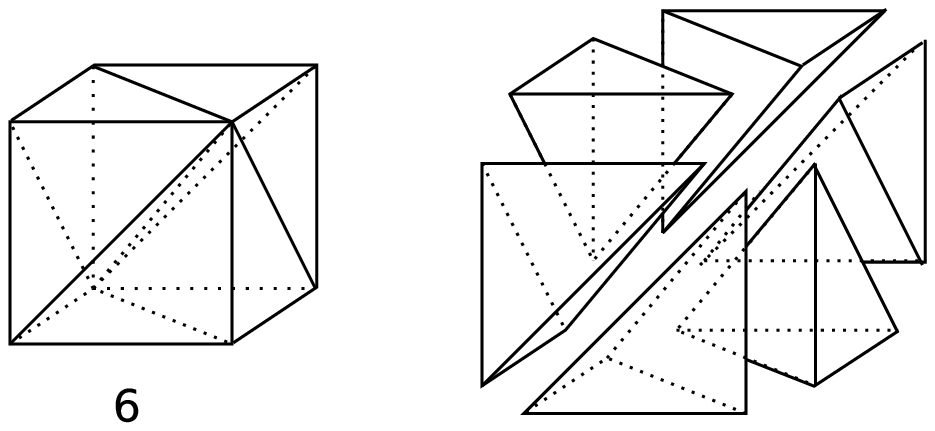}
  \caption{
    Middle cut triangulation and Kuhn's triangulation: dividing a cube into five or
    six tetrahedra, respectively.
  }
  \label{F:pentasection}
  \end{center}
  \end{figure}

%-----------------------------------------------------------------------------
{\bf Cell triangulation.}
For a cubical cell, there exist two possible triangulations
(see figure \ref{F:pentasection}): the cell can be divided into either
five or six tetrahedra. The former is known as \emph{middle cut
triangulation}~\cite{Sallee:1984:MCT}, and the latter as \emph{Kuhn's 
triangulation}~\cite{Kuhn60:CLT}. While Kuhn's triangulation produces simplices
of equal shape and can be more easily extended to an arbitrary number of
dimensions~\cite{CMin03:Simplicial}, the middle cut triangulation produces
higher quality tetrahedra, in the sense that their angles are less acute than
in a Kuhn triangulation, which produces to a finite element matrix system with 
better condition number.
Usually, the quality of tetrahedra is described by the ratio $R/3r$ of radii of
circumscribed ($R$) to inscribed ($r$) spheres. For a middle cut triangulation 
the central tetrahedron has, obviously, minimum possible aspect ratio $R/3r =
1$, while for corner tetrahedra we have $R/3r =\frac{1+\sqrt{3}}{2} \approx
1.37$. At the same time, in Kuhn's triangulation all tetrahedra have aspect
ratio $1+1/\sqrt{3} \approx 1.58$.

% /---------------------------------------------------------------------\
% \---------------------------------------------------------------------/

\begin{figure}[!htp]
  \begin{center}
  \includegraphics[width=0.3\textwidth]{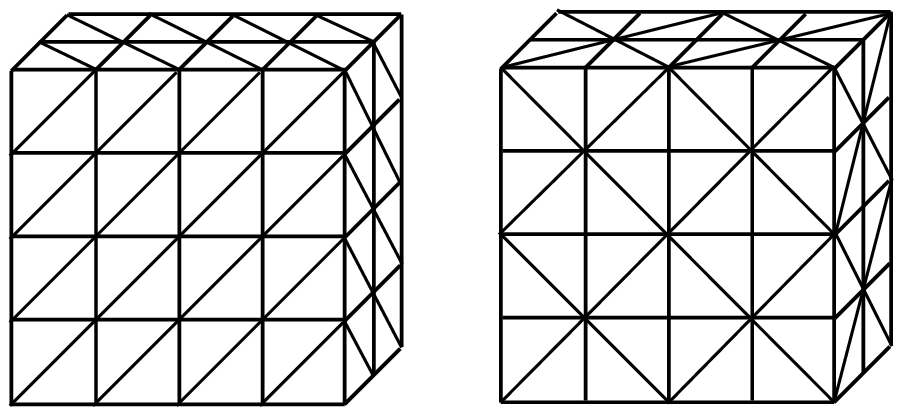}
  \caption{
    Uniform (left) and clustered (right) block triangulations. With 
    Kuhn's triangulation of each cell, both types of block triangulation
    can be done in conforming way for neighboring cells. Cells,
    triangulated by middle cut, can only be arranged into clustered
    configuration.
  }
  \label{F:block_triangulations}
  \end{center}
  \end{figure}

%-----------------------------------------------------------------------------
{\bf Block triangulation.}
A block can be triangulated in many possible ways. We will consider the two most
straightforward and symmetric ones, referring to them as \emph{uniform} and
\emph{clustered} block triangulations. Both types of triangulation are
locally symmetric with respect to local space inversion at any inner
vertex of the mesh.
In \emph{uniform} block triangulation, exactly the same cell triangulation,
with the same orientation, is applied to all cells in a block
(see figure~\ref{F:block_triangulations}). This produces a conforming mesh for
Kuhn's triangulation, but fails to produce a conforming one for the middle cut one,
since the triangulation patterns between neighboring cell interfaces are not
compatible.

% /---------------------------------------------------------------------\
% \---------------------------------------------------------------------/

% /---------------------------------------------------------------------\
  \begin{figure}[!htp]
  \begin{center}
  \includegraphics[height=0.1\textheight]{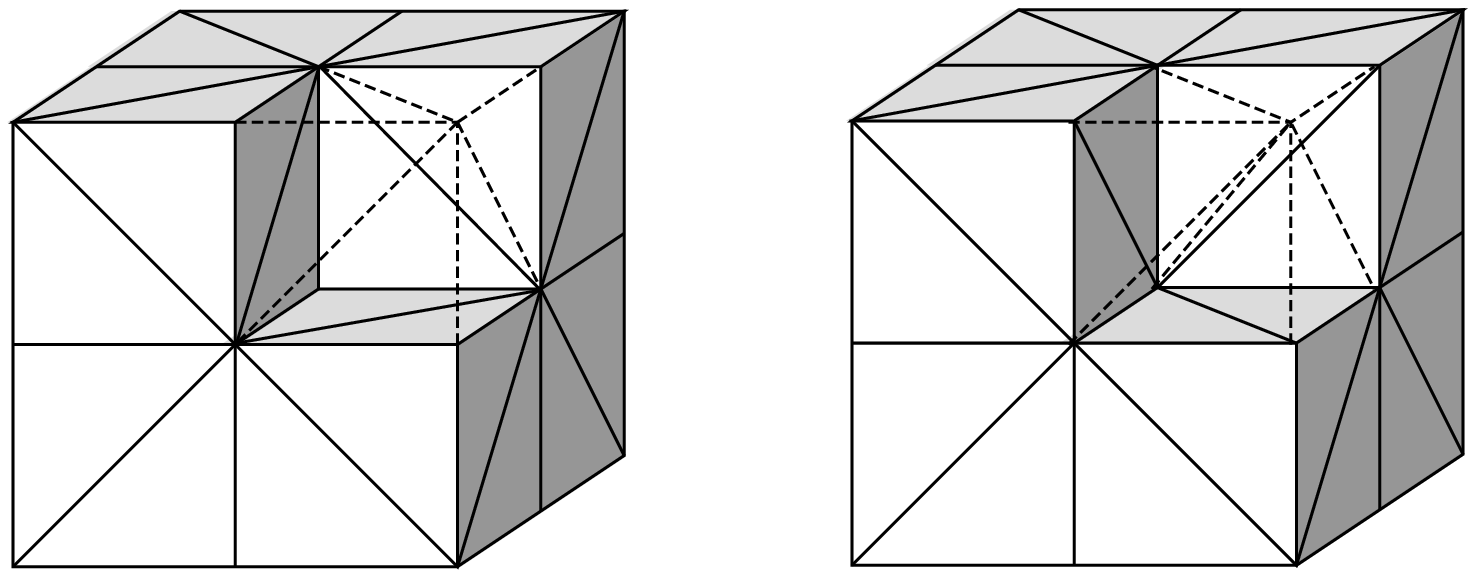}
  \caption{
    Union jack triangulation of $2\times 2\times 2$-cell cluster. The
    triangulation pattern on each side of the cluster is the same.
    Inside, each cluster can have either middle cut (left) or Kuhn 
    (right) triangulations.
  }
  \label{F:central_triangulation}
  \end{center}
  \end{figure}
% \---------------------------------------------------------------------/

To handle this compatibility problem, it is convenient to group neighboring
cells into $2\times 2 \times 2$ clusters and triangulate each cluster as shown in figure
~\ref{F:central_triangulation}, with "union jack" patterns on each side. This
guarantees conformity between neighboring clusters. We will label this type of
triangulation as \emph{clustered} block triangulation. It can be used with
both types of cell triangulation.

\begin{figure}[!htp]
  \begin{center}
  \includegraphics[height=0.2\textheight]{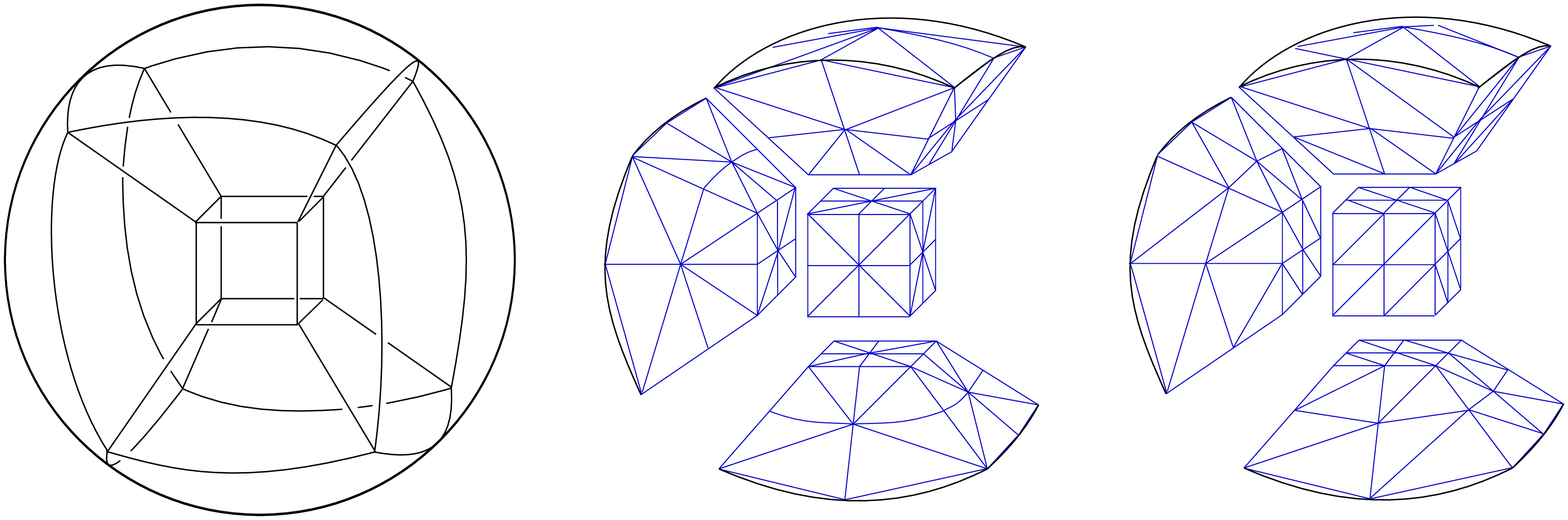}
  \caption{
    Seven-block system for the sphere (left), and its triangulations,
    generated using clustered (center) and uniform (right) block
    triangulations.
  }
  \label{F:sevenpatch}
  \end{center}
  \end{figure}

%-----------------------------------------------------------------------------
{\bf Multi-Block triangulation.}
In a multi-block system, both uniform and clustered triangulations of each
block can be arranged into a conforming simplicial mesh. Obviously, clustered
triangulations of each block with even number of cells on the interfaces
between the blocks assemble themselves in a conforming manner. A uniform
triangulations arrangement can be constructed from clustered
triangulation of the same system by first removing and then replicating
layers of cells with Kuhn's triangulation.
Two different types of semi-structured multi-block triangulations, used in
this paper, are shown in figure~\ref{F:sevenpatch}.

% /---------------------------------------------------------------------\
% \---------------------------------------------------------------------/

%=============================================================================
\section{Quality of our finite element solutions on semi-structured multi-block triangulations}
\label{S:QualSol}
%-----------------------------------------------------------------------------
In this section we perform a numerical study of the quality of our finite element
solutions obtained using semi-structured multi-block triangulations. We
investigate not only the solution itself, but also answer the question
of whether this solution is appropriate for finite-difference evolution
codes with high-order numerical derivative operators. After introducing
the domain structure and weak formulation of the second-order elliptic
equation, we evaluate the solution convergence order, and show
superconvergence for quadratic finite elements. Then we evaluate the finite 
element solution at the multi-block gridpoints, apply various
high-order finite difference operators and check the convergence orders of
its first and second numerical derivatives. The observed convergence orders
are consistent with the expected values. 

%-----------------------------------------------------------------------------
\subsection{Domain structures}
\label{SS:DomStr}
\begin{figure}[!htp]
  \begin{center}
  \begin{tabular}{ccc}
  \includegraphics[width=0.25\textwidth]{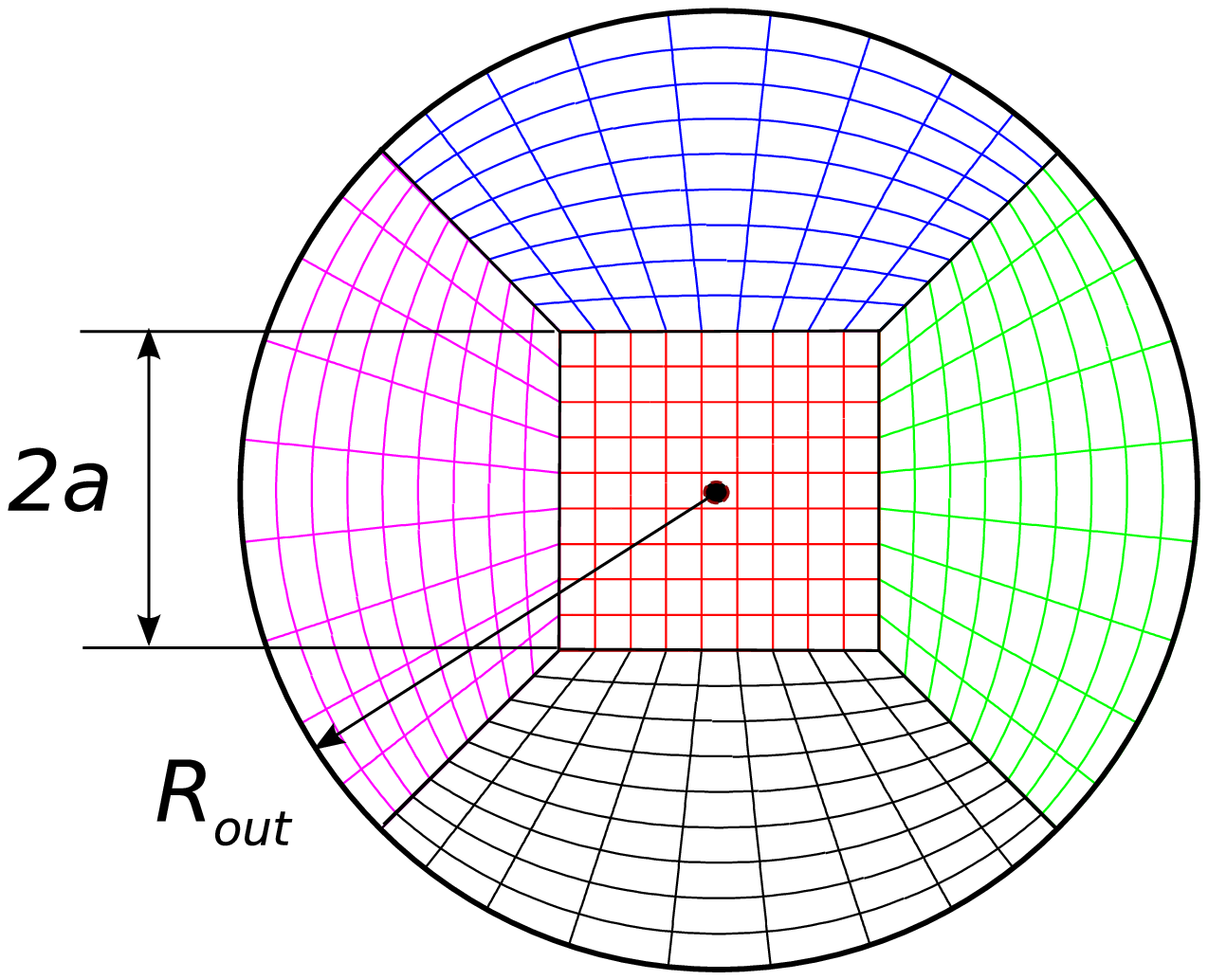} &
  \includegraphics[width=0.27\textwidth]{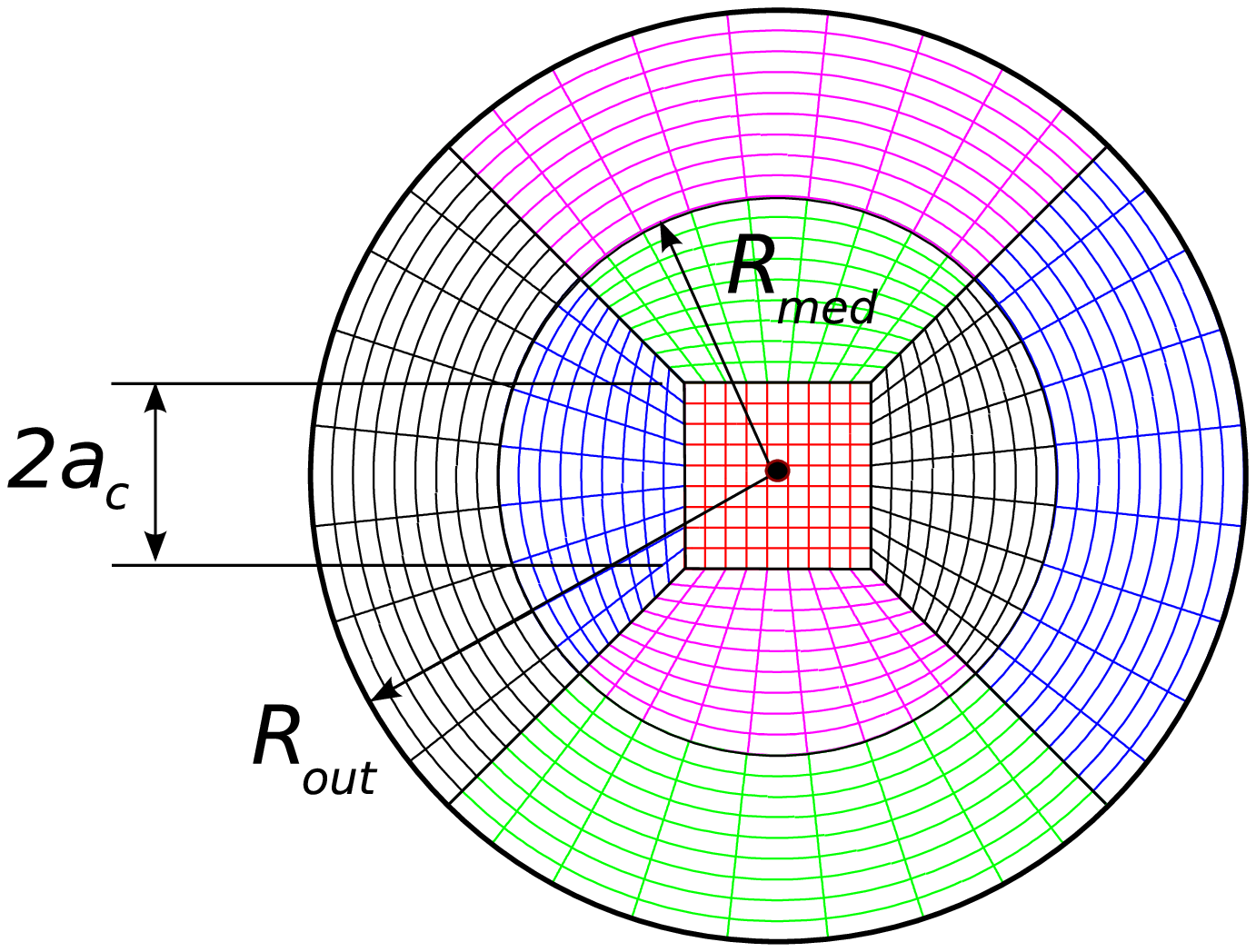} &
  \includegraphics[width=0.20\textwidth]{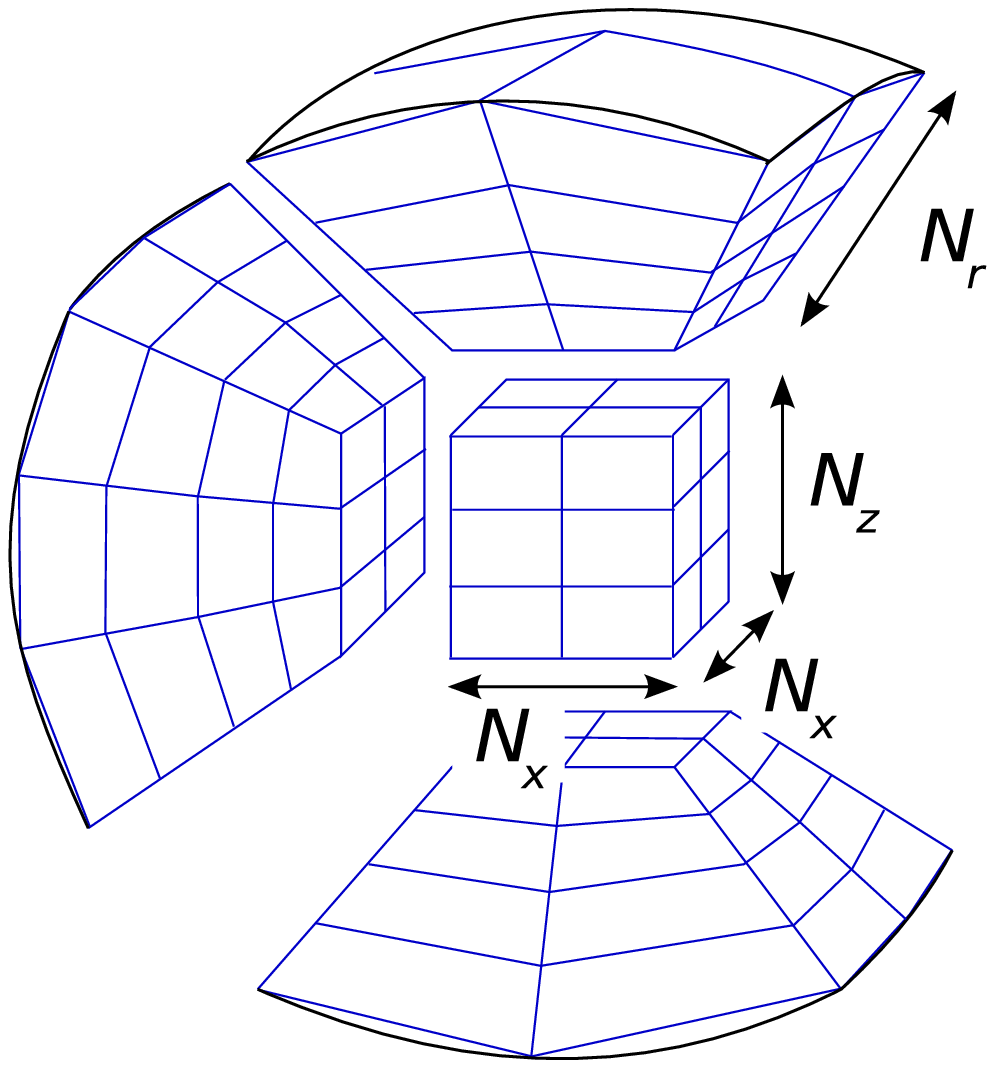} \\
  {\tiny (a)} & {\tiny (b)} & {\tiny (c)}
  \end{tabular}

  \caption{
    An equatorial cut of (a) seven-block and (b) thirteen-block systems.
    (c) Grid dimensions for the seven-block system.
  }
  \label{F:patchsystems}
  \end{center}
  \end{figure}

%-----------------------------------------------------------------------------
Both \FETK{} and QUILT were developed to handle equations on general manifolds, 
with an arbitrary number of charts. However, for the domain structures 
considered in this paper we can embed the 
computational domain into a reference Euclidean 3-dimensional space with a single fixed 
Cartesian system of coordinates to label all vertices and nodes of the mesh, 
and we do so. 
The domain of interest here will then be a spherical domain of radius $R$, equipped with
a seven-blocks or thirteen-blocks system (see figure~\ref{F:patchsystems}),
with local patch coordinate transformations defined as 
in~\cite{diener-2007-32}, \cite{lehner-2005-22}. 
% /---------------------------------------------------------------------\
% \---------------------------------------------------------------------/
The seven-block geometry (see figure \ref{F:patchsystems}) is fully specified
by fixing the outer sphere radius $R_{out}$ and the side of the inner cubical
patch $a_c$. In the innermost, cubical patch, there are $N_x\times N_y\times
N_z$ points, where $N_x=N_y=N_z:=N$. Since the grids are conforming, in
the six blocks surrounding the inner one there are $N\times N \times N_r$
points. 
The thirteen-block geometry, in turn, can be seen as a seven-block one of
radius $R_{med}$  and $N\times N\times N_{r,in}$ points, surrounded by six
additional blocks with $N\times N\times N_{r,out}$ points each, with a total
radius $R_{out}$. The advantage of the thirteen-block setup is that the
transversal grid layers in the outermost system are perfectly spherical
surfaces. These are very convenient in certain applications which require
integration over such surfaces, as in the the study of the multipole structure
of radiation in the wave zone (see, for example, \cite{Pazos:2006kz}), since no interpolation is needed for those
integrations (resulting in both higher accuracy and speed).
Throughout this paper, when we change the resolution we keep the ratios
$N:N_r$ and $N:N_{r,in}:N_{r,out}$ in the seven- and thirteen-block
systems, respectively, fixed. 
Multiple domains with different values of $N$ produce a sequence of domain
triangulations \{$\mathcal{T}_h$\} with maximal simplex diameters $h$
inversely proportional to $N$. Therefore, it is convenient to use $N$ as a
scaling factor in convergence tests, and we do so. 

%-----------------------------------------------------------------------------
\subsection{Second-order elliptic equation and its weak form}
\label{SS:SecOrd}
%-----------------------------------------------------------------------------
Let $S_R$ represent our spherical domain with radius $R$, centered at the
origin, and let $\pd S_R$ be its outer boundary. The equations of interest in
this paper are of the form:

\begin{eqnarray}
\label{Eq:BrillWave1}
-\nabla^2\psi(\bar{x}) + V(\bar{x})\psi(\bar{x}) & = & 0 ~\text{~in~} S_R, \\
\psi(\bar{x}) & = & \psi_D ~\text{~on~} \pd S_R
\end{eqnarray}

where $\psi_D$ is a Dirichlet boundary condition, and $\nabla^2$ is the Laplace
operator. We only consider the case when both the potential $V$ and the
boundary conditions $\psi_D$ are axisymmetric.  Following the weak formulation
example above (section ~\ref{SS:WeakForm}), we obtain the nonlinear
weak form and the bilinear linearization form:
\begin{eqnarray}
\langle F(\psi),\phi \rangle 
  & = &\int_{S_R} \left(\nabla\psi \cdot \nabla\phi
    +  \psi V\phi \right) ~dx \\
\langle DF(\psi)\chi,\phi \rangle 
  & = &\int_{S_R} \left(\nabla\chi \cdot \nabla\phi
    +  \chi V\phi \right) ~dx 
\end{eqnarray}

with $\psi(\bar{x}) \in \bar{\psi} + H^1_0(S_R)$, 
and $\phi(\bar{x})~,~\chi(\bar{x}) \in H^1_0(S_R)$ as discussed in
section~\ref{S:SolvFEtk}. These two forms and the Dirichlet boundary
function are everything we need to specify the problem in~\FETK{}.

We also consider the same problem with more complex Robin boundary conditions: 
\begin{equation}
\label{Eq:RobinBC}
\pd_r\psi(\bar{x}) = \frac{1-\psi(\bar{x})}{r} ~\text{~on~} \pd S_R \, ,
\end{equation}
which has the following nonlinear weak and bilinear linearization forms,
\begin{eqnarray}
\label{Eq:WeakFormRobinBC}
\langle F(\psi),\phi \rangle 
  & = &\int_{S_R} \left(\nabla\psi \cdot \nabla\phi
    + \psi V\phi \right) ~dx 
    + \frac{1}{R} \int_{\pd S_R} (\psi-1)\phi ~ds \\
\label{Eq:BilinearLinearizationRobinBC}
\langle DF(\psi)\chi,\phi \rangle 
  & = &\int_{S_R} \left(\nabla\chi \cdot \nabla\phi
    + \chi V\phi \right) ~dx 
    + \frac{1}{R} \int_{\pd S_R} \chi ~ \phi ~ds 
\end{eqnarray}

where now $\psi(\bar{x}),~\phi(\bar{x}),~\chi(\bar{x}) ~\in~ H^1(S_R)$.
We will use these Robin boundary conditions when solving for Brill waves below 
in section~\ref{S:BrillWaveID}.

%-----------------------------------------------------------------------------
\subsection{Testing convergence of the solution}
\label{SS:SecOrdConvSol}
% - - - - - - - - - - - - - - - - - - - - - - - - - - - - - - - - - - - - - - 

As a test problem for our approach to solving elliptic equations on a
semi-structured grid using finite elements, we solve
equation~\eqref{Eq:BrillWave1} with three different potentials: 

\begin{eqnarray}
&V_A    &= -3\omega^2,                                 \notag \\
&V_B(r) &= \frac{2(r^2-3r_0^2)}{(r^2+r_0^2)^2},        \notag \\
&V_C(\rho,z) &=
    -\frac{2}{Z}\left(1 + \frac{4z^2}{R}
                          (S+\frac{\sigma_r r_0^2}{Z})\right)
    +\frac{2(2C+3S)}{R} + 2\left(\frac{2rS}{R}\right)^2,
    \notag
\end{eqnarray}

where
  $R=r^2-\sigma_r r_0^2\cosh{\frac{r^2}{r_0^2}}$, 
  $Z=z^2+\sigma_z^2$,
  $C=\sigma_r\cosh{\frac{r^2}{r_0^2}}$,
  $S=\sigma_r\sinh{\frac{r^2}{r_0^2}} - 1$.

These are such that they produce the following solutions:
\begin{itemize}
\item[(A)] Plane wave: 
           $\psi_A(x,y,z)~=~\cos(\omega x)\cos(\omega y)\cos(\omega z)$
\item[(B)] Spherically-symmetric pulse with width $r_0$, concentrated at the
           origin, falling off with order $1/r^2$ as $r\to\infty$:
           $\psi_B(r)~=~\frac{1}{1 + r^2/r_0^2}$
\item[(C)] Toroidal solution, with radius $\sim r_0$ and width $\sim\sigma_r$
           in the radial direction and $\sigma_z$ in the vertical one:
           $\psi_C(\rho,z)=
             (
              \cosh{(\frac{r^2}{r_0^2})}-
              \frac{r^2}{\sigma_r r_0^2}
             )^{-1}
             (1 ~+~ z^2/\sigma_z^2)^{-1}$
\end{itemize}

% /---------------------------------------------------------------------\
  \begin{figure}[!htp]
  \begin{center}
  \begin{tabular}{cc}
  \includegraphics[width=0.45\textwidth]{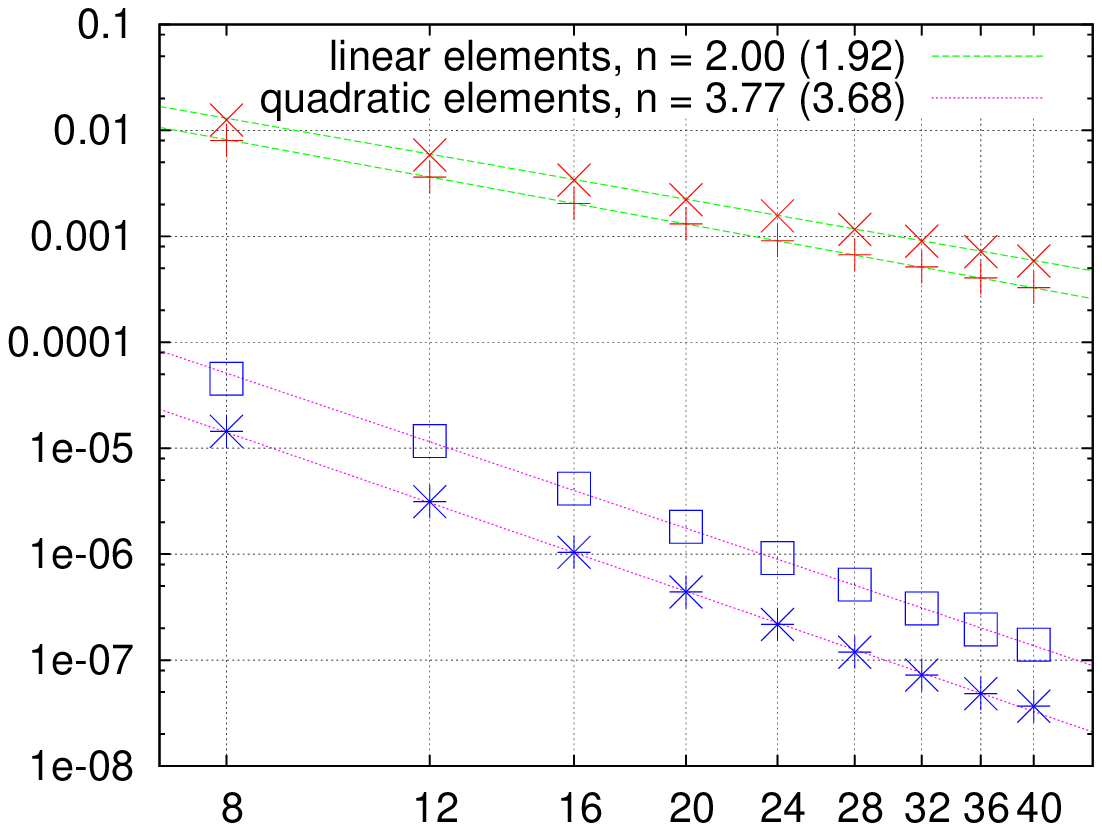} & 
  \includegraphics[width=0.45\textwidth]{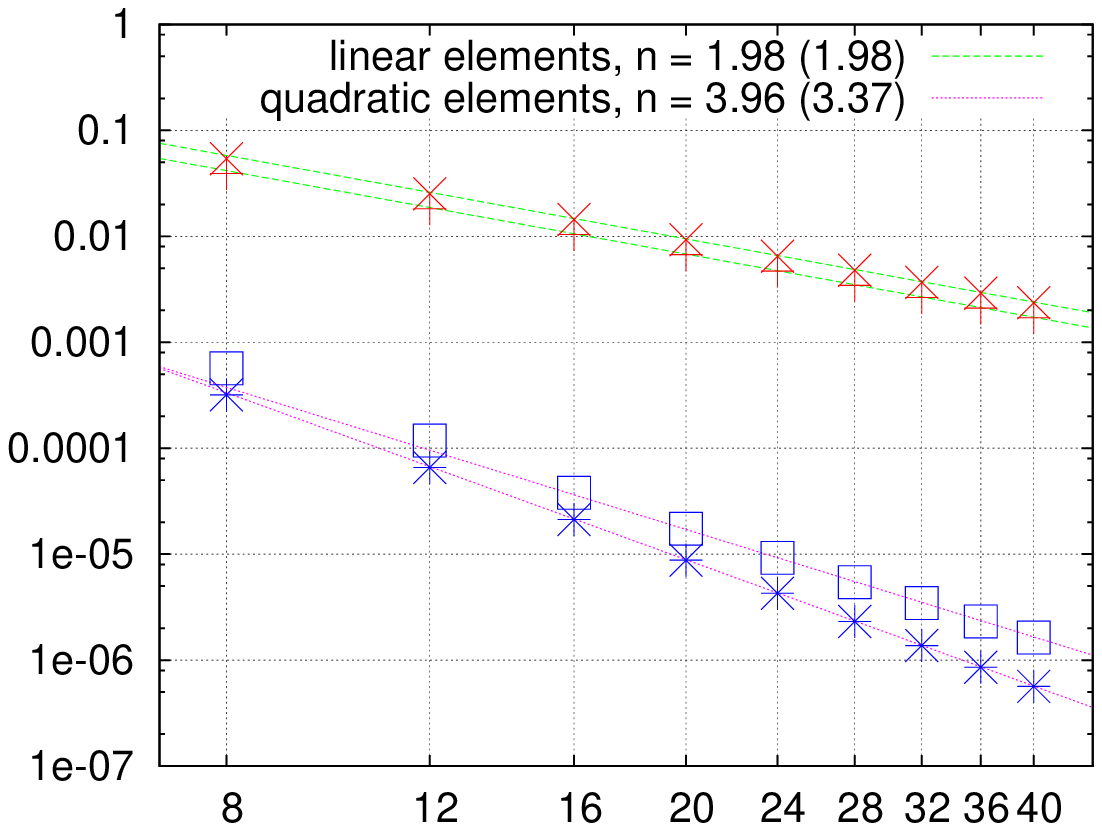} \\
  {\tiny (a)} & {\tiny (b)}\\
  \end{tabular}
  \includegraphics[width=0.45\textwidth]{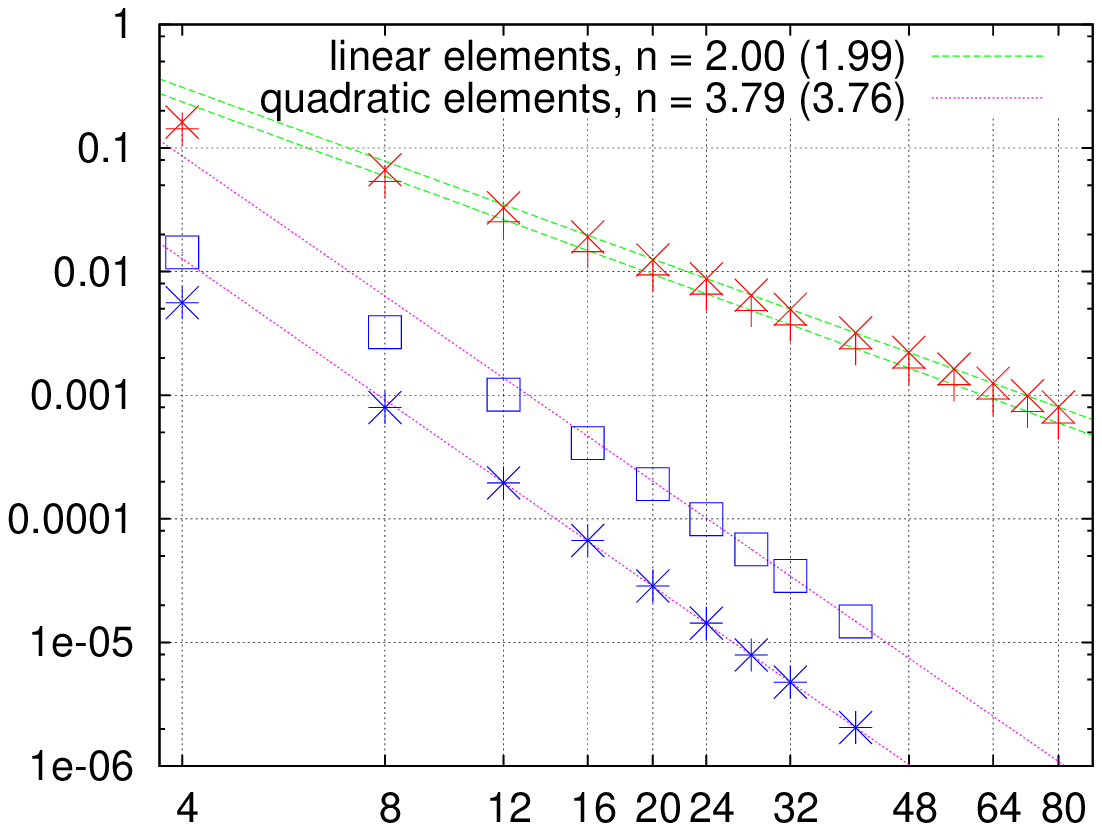} \\
  {\tiny (c)}
  \caption{
    Convergence of the solution error, in the $l_2$ and $l_{\infty}$
    norms, with respect to the number of points $N$, for linear and
    quadratic finite elements. The plots (a), (b) and (c) correspond to
    the potentials $V_A$, $V_B(r)$ and $V_C(\rho,z)$, respectively. From
    each pair of lines, the upper one represents the $l_{\infty}$ norm,
    and the lower one the $l_2$ norm. 
    Each plot shows the convergence orders $n$, obtained by linear fit
    using the four points with highest resolution. The numbers in
    brackets give the convergence orders in the $l_{\infty}$ norm.
  }
  \label{F:err-nx}
  \end{center}
  \end{figure}
% \---------------------------------------------------------------------/

These three potentials are chosen such that the solution for $\psi$ is known in
closed form, and they all differ in the way they are adapted to the underlying
multi-block grid.
The first potential, $V_A$, represents a simple periodic wave, with the
mesh not adapted to the shape of the wave. Second potential is supposed to
model a situation where the wave is concentrated near the origin, in the
central cubical patch of the "cubed sphere" domain. In this case, the grid
resolution is adapted to the solution, but not the coordinate lines. Finally,
for the potential $V_C$ (which has toroidal shape), both the resolution (in
the $\theta$ and $r$ directions), and coordinate lines (in $\ji$ direction)
are adapted to the solution.

All the test problems were solved on the same 7-patch spherical domain, with
dimensions $R_{out}~=~10$, $a_c~=~2.5$, and fixed grid size ratios
$N~:~N_r~=~1~:~2$.  The test problems used the following set of
parameters: for $V_A$: $\omega = 0.1$, for $V_B(r)$: $r_0 = 4$, and for
$V_C(\rho,z)$: $r_0=8$, $\sigma_r=1.2$, $\sigma_z=4$.

It is well-known (see, for example, \cite{R:Bra07}, \cite{R:BrS03}, 
\cite{R:eg04}) that in case of the optimal approximation, the convergence rate
of continuum-level error norms, $||u_h-u_e||_2$ and $||u_h-u_e||_{\infty}$,
defined in a usual manner,

\begin{eqnarray}
  ||u_h-u_e||_2 &=& \left(\int_{\Mfold}(u_h(x)-u_e(x))^2 dx\right)^{1/2}, \\
  ||u_h-u_e||_{\infty} &=& \max_{x\in\Mfold}|u_h(x)-u_e(x)|
\end{eqnarray}

for the standard uniform refinement, is determined by the approximation power
of the finite element function spaces (here, $u_e(x)$ is the exact solution,
$u_h(x)$ is its finite element approximation, and index $h$ denotes the
maximum simplex diameter in the domain triangulation $\mathcal{T}_h$). If
piecewise polynomials of fixed order $p$ are used, then the order of
convergence of the continuum-level error norms is $p+1$.

In order to measure the error of the grid solutions, we use discrete $l_2$
and $l_{\infty}$ norms, sampled at the nodes and normalized by the
corresponding norm of $u_e$:

\begin{align}
&\epsilon_{h,2} = 
\frac{\left(\frac{1}{N}\sum_{k=1}^N (u_{h,k}-u_e(x_k))^2 \right)^{1/2}}
     {\left(\frac{1}{N}\sum_{k=1}^N u_e(x_k)^2 \right)^{1/2}}
&\epsilon_{h,\infty} = 
\frac{\max_{k} |u_{h,k}-u_e(x_k)|}
     {\max_{k} |u_e(x_k)|}
\notag
\end{align}

The plots on figure~\ref{F:err-nx} show the convergence of $\epsilon_{h,2}$ and
$\epsilon_{h,\infty}$ with $N$ ($N$ is inversely proportional to the maximum
mesh diameter $h$, see section~\ref{SS:DomStr}). Note that the convergence
orders in  the $l_2$ and
$l_{\infty}$ norms agree, which means that the pointwise convergence order is
the same everywhere. The observed convergence order for linear elements is
$2$, which is supposed to be the case, since the order of piecewise
polynomials is odd. For quadratics, we expect to have superconvergence, and
indeed, the observed convergence rate is $\approx 4$.

%-----------------------------------------------------------------------------
\subsection{Testing convergence of numerical derivatives of the solution}
\label{SS:SecOrdConvND}
% - - - - - - - - - - - - - - - - - - - - - - - - - - - - - - - - - - - - - - 
To set up initial data for our General Relativity evolution codes, we need not only the solution
itself but also its first spatial derivatives, because we use a first-order formulation
of the Einstein evolution equations. In total, our finite element solution has to be
differentiated twice: once, when setting initial data, and one more time when
computing the evolution equations. In this subsection we numerically study 
how our obtained finite element solutions behave under two numerical
finite-difference differentiations in terms of convergence.

% /---------------------------------------------------------------------\
  \begin{figure}[htbp]
  \begin{center}
  \includegraphics[width=0.40\textwidth]{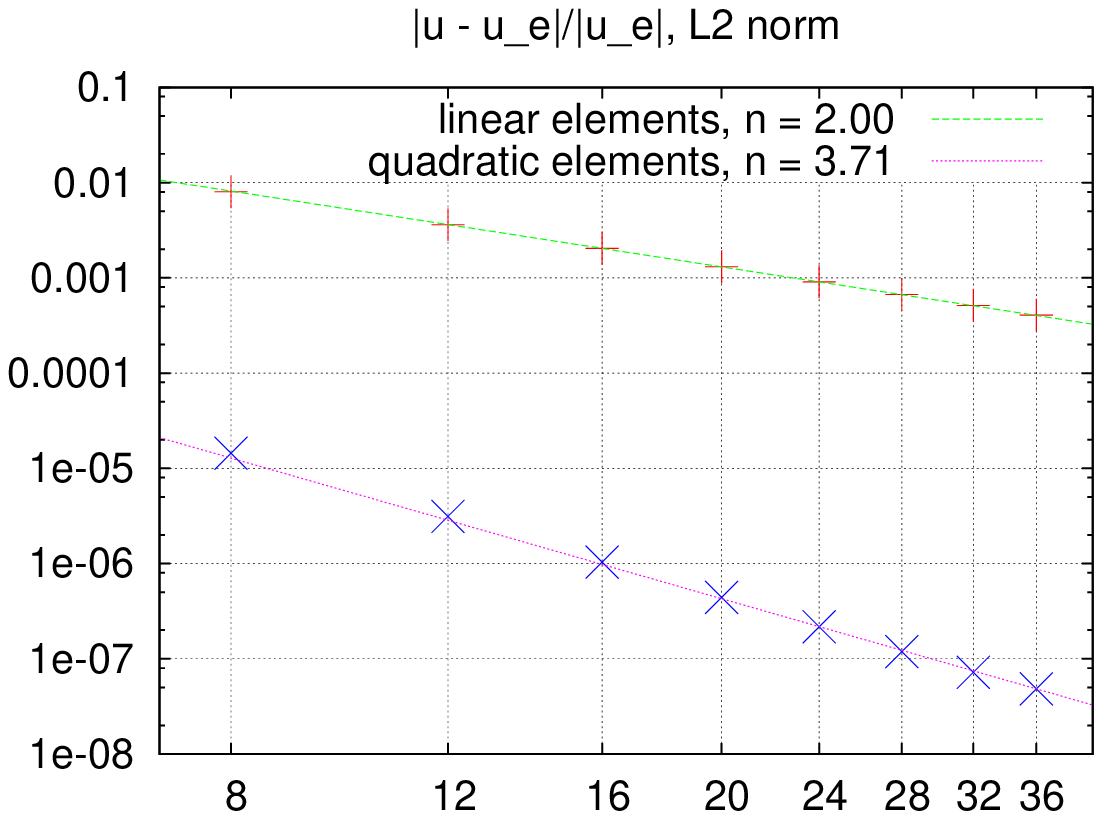}\\
  {\tiny (a)}\\
  \begin{tabular}{cc}
  \includegraphics[width=0.35\textwidth]{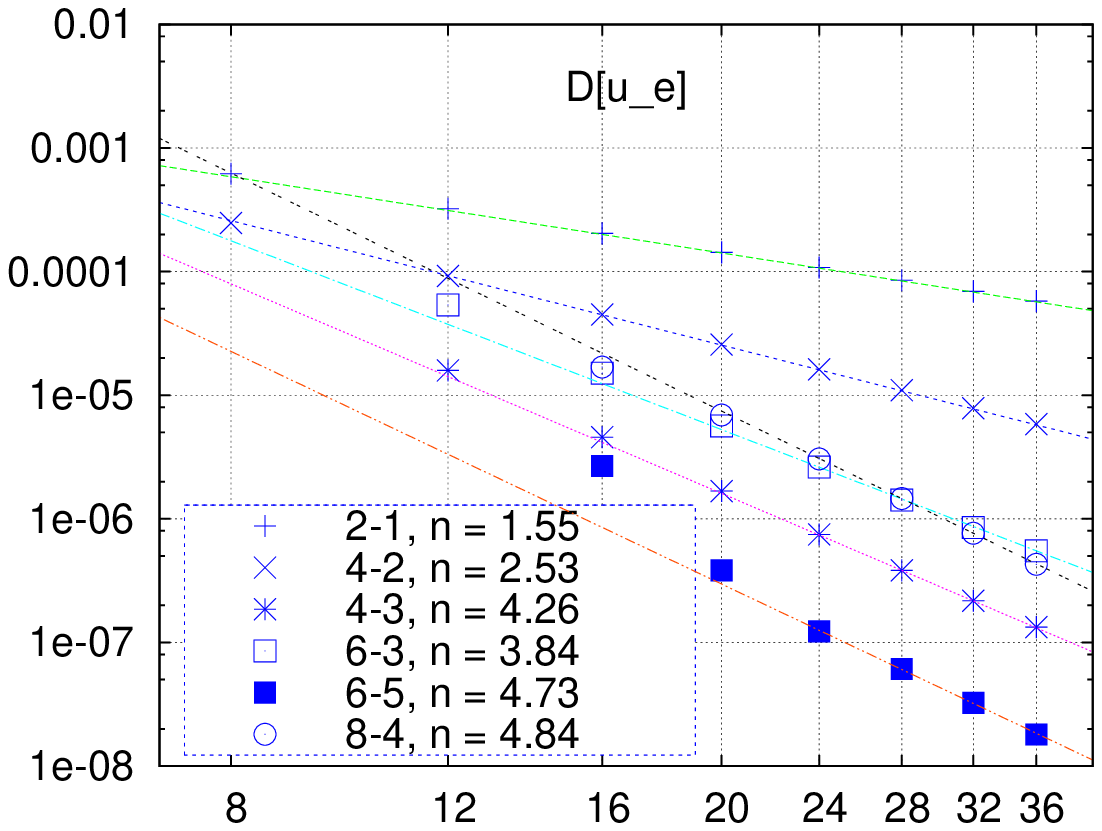} &
  \includegraphics[width=0.35\textwidth]{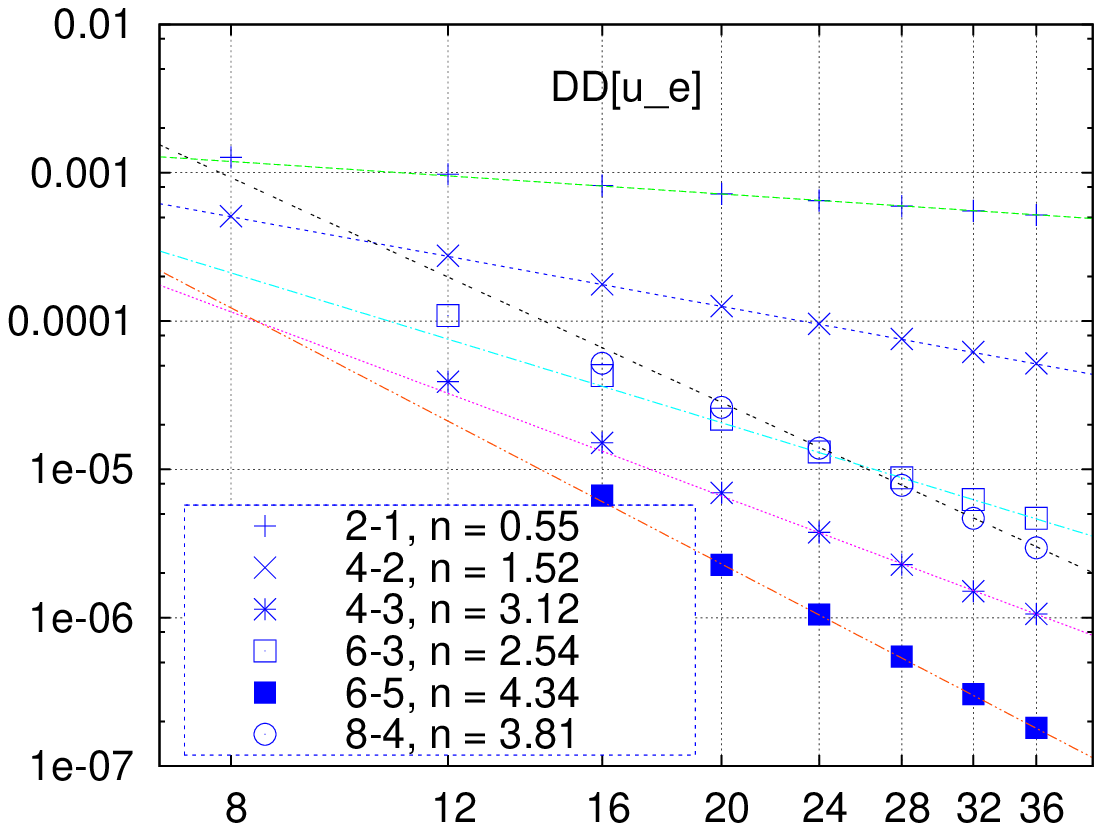} 
  \end{tabular}\\
  {\tiny (b)} \\
  \begin{tabular}{cc}
  \includegraphics[width=0.35\textwidth]{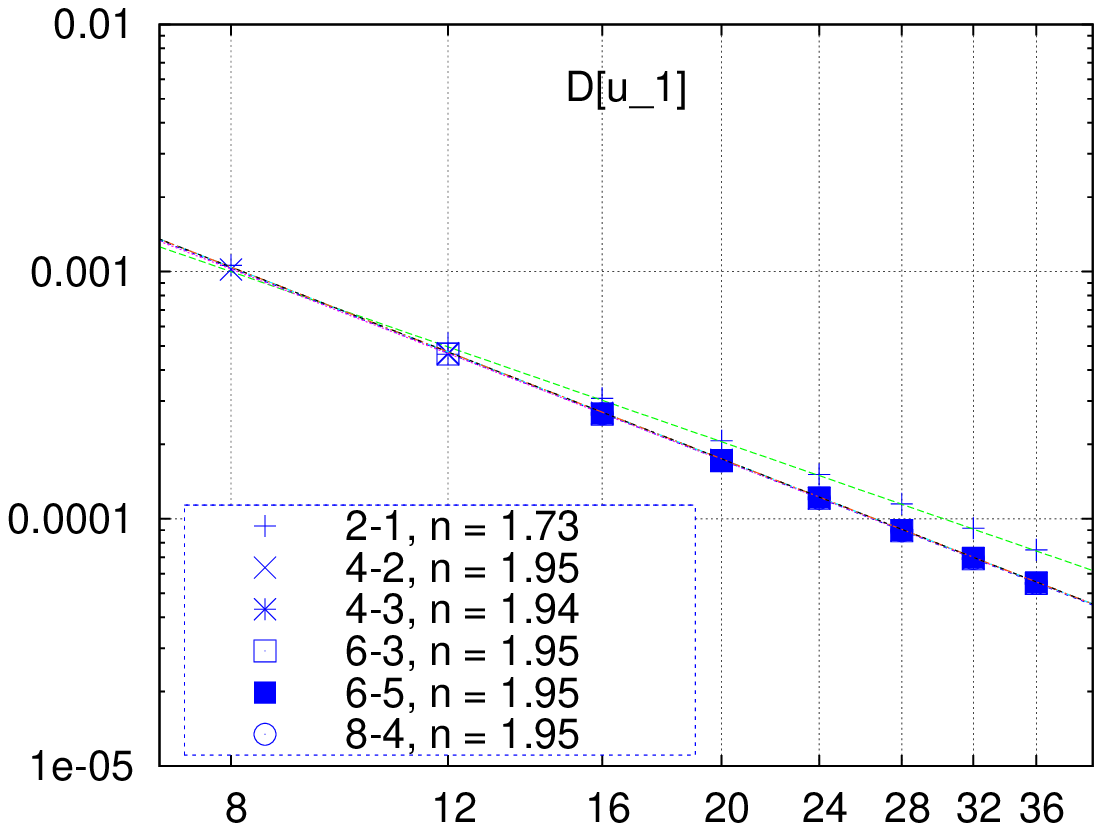} &
  \includegraphics[width=0.35\textwidth]{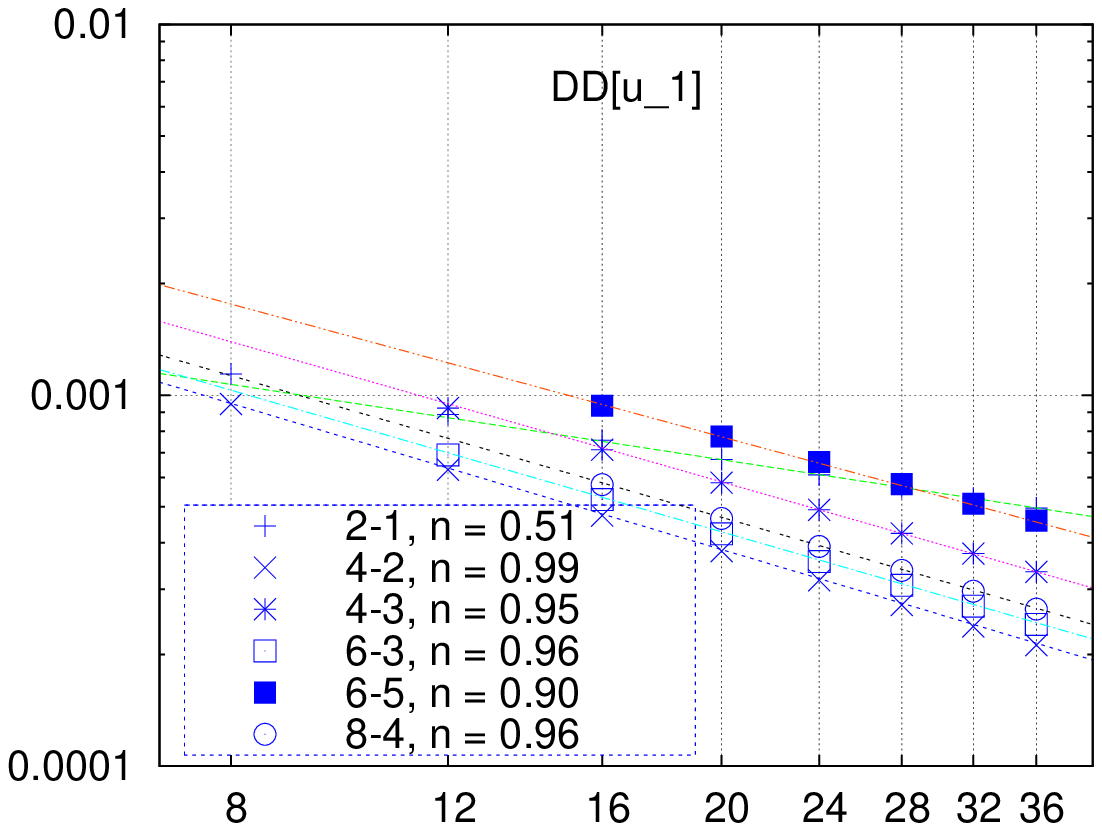} 
  \end{tabular}\\
  {\tiny (c)}\\
  \begin{tabular}{cc}
  \includegraphics[width=0.35\textwidth]{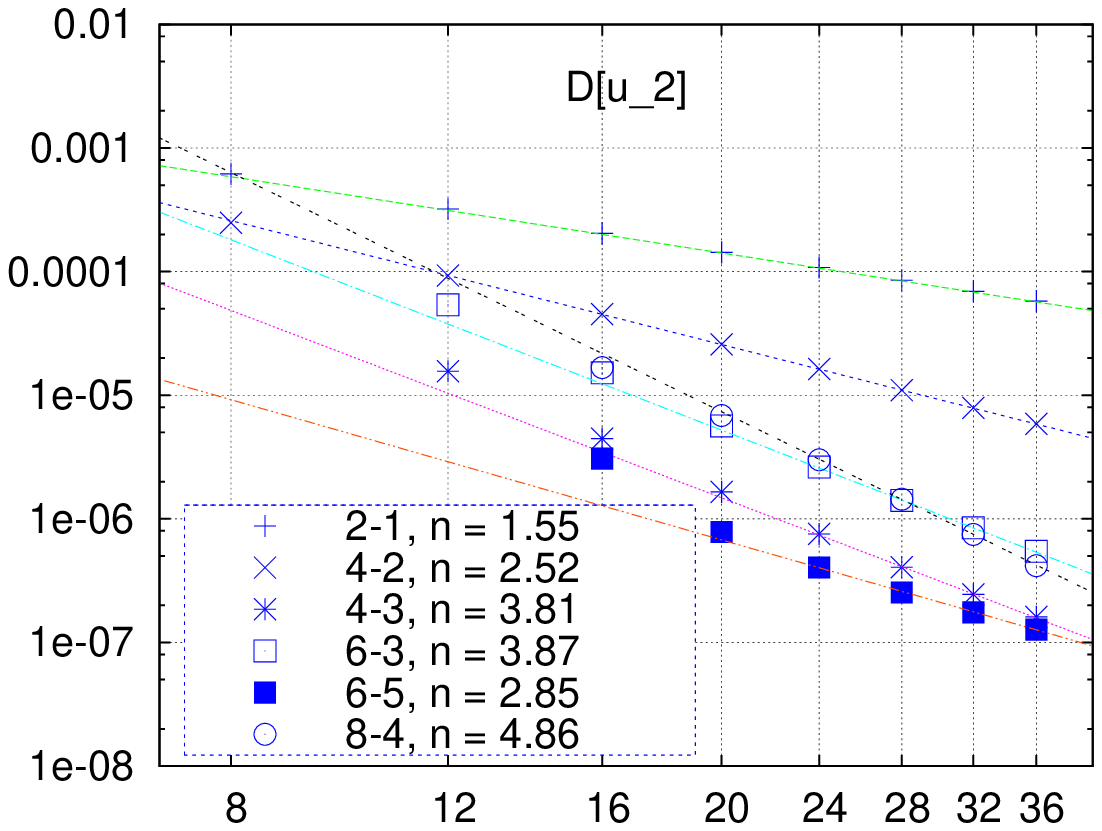} &
  \includegraphics[width=0.35\textwidth]{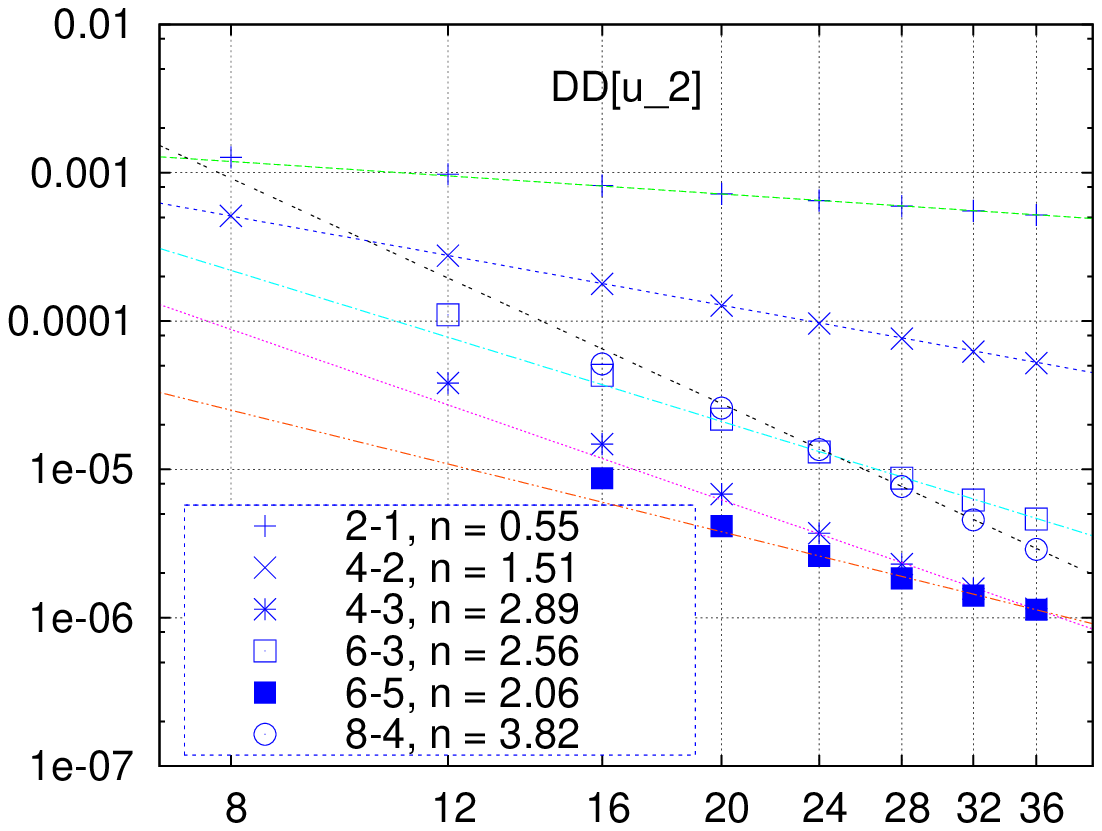}
  \end{tabular}\\
  {\tiny (d)}\\
  \caption{
    (a) Convergence of the solution error in the $l_2$-norm, for linear and
    quadratic finite elements, for the test problem with potential $V_A$ and 
    $\omega=0.1$. 
    (b,c,d) Convergence in the $l_2$-norm of the first (left) and second (right)
    numerical FD derivatives of: (b) the exact solution, restricted to the FD grid;
    (c) the numerical solution obtained using linear finite elements; (d) the
    numerical solution  obtained using quadratic finite elements. In all cases
    several FD operators are used to compute the derivatives, and the
    resulting convergence factors (denoted by $n$) are shown
  }
  \label{F:cos3-ndx-norm2}
  \end{center}
  \end{figure}
% \---------------------------------------------------------------------/

If we were using completely unstructured meshes and interpolated the finite
element solution to the multi-block grids used in our evolutions, the resulting grid solution
would have an error of order $O(h^{p+1})$. Two numerical differentiations in
this case would take away two orders of convergence, leading to an
unacceptable low convergence order.
However, as discussed above, when using semi-structured grids special conditions which lead to
superconvergence can be met, and the convergence rate of numerical derivatives
improve. 

The procedure for converting the finite element solution into the grid solution is
trivial: the value of the solution $u(x)=\sum_{i=1}^N c_i\phi_i(x)$ at a node $i$
is simply the corresponding nodal coefficient $c_i$. In order for the nodes to
coincide with the curvilinear gridpoint in the blocks, we restrict our finite element
solutions only to vertex nodes, and omit other types of nodes, such as
mid-edge ones.

In our multi-block evolutions we use new, efficient high-order
finite differencing (FD) operators satisfying the summation-by-parts (SBP)
property, constructed and described in detail in~\cite{diener-2007-32}. If $D$ is a
one-dimensional differential operator, the SBP property means that for any
grid functions $u$ and $v$ on a segment $[a,b]$ with a constant grid spacing
$h$, the following condition is satisfied

$$
\langle Du,v\rangle + \langle u,Dv\rangle = u(b)v(b) - u(a)v(a),
$$

where $\langle\cdot,\cdot\rangle$ denotes the scalar product between two grid
functions, defined by the SBP matrix $\Sigma = ||\sigma_{ij}||$ associated
with $D$:

$$
\langle u,v\rangle = \sum_{i,j} \sigma_{ij} u_i v_j
$$

In this paper, we use the following FD operators: $D_{2-1}$, $D_{4-2}$,
$D_{4-3}$, $D_{6-3}$, $D_{6-5}$ and $D_{8-4}$. The pair of numbers in the FD
operator's subindex reflects the convergence order at interior points and at
 points at and close to the boundary. In more detail: for a generic FD
 operator $D_{a-b}$, the convergence order in the interior is $a$ and at and
 close to the boundaries it is $b\le a$. The convergence of the 
numerical derivative in the $l_2$ norm is at least of order $b+1$
\cite{Gustafsson98:BC}
and in the $l_{\infty}$ norm -- it is at least of order $b$.

The operators $D_{2-1}$, $D_{4-2}$, $D_{6-3}$ and $D_{8-4}$ are diagonal norm
(scalar product) based, as they satisfy SBP with respect to certain diagonal
scalar product (norm) of grid functions. $D_{4-3}$ and $D_{6-5}$ are so-called
"full restricted norm" operators. They satisfy SBP with respect to norms which
are not necessarily diagonal, but only restricted to be diagonal at the boundary
points. The diagonal norm FD operators have several advantages compared to
full restricted norm ones in terms of stability properties. However, 
they exhibit lower order of convergence at and close to 
boundary points. While the full restricted norm operators are only one order
less convergent at the boundary, diagonal norm operators lose half the 
convergence order compared to the interior points.

After the finite element solution is converted to a grid function and then
numerically differentiated with FD, the resulting convergence factor for the
numerical derivatives depends not only on the order of the finite element basis
polynomials, but also on the order of the FD operator.  In general one expects
that FD operators of sufficiently high order will preserve the original
convergence of the finite element solution. This is a very nontrivial
mathematical result in superconvergence theory (see Section 8.2
of~\cite{Wahlbin95:superconvergence} and references therein). In this section we
test it for first and second FD derivatives (where by 'second FD derivative' we
mean 'first FD derivative, applied two times', as opposed to a second-order FD
derivative). The second FD derivative is expected be one order less convergent.

% +----------------+----------------+----------------+----------------+----------------+----------------+
  \begin{table}
  \begin{center}
  {\small
  \begin{tabular}{||c|c|c||}
   \hline
   potential   & linear & quadratics \\ \hline
   $V_A$       & 2.00    & 3.71       \\ \hline
   $V_B$       & 1.98    & 3.96       \\ \hline
   $V_C$       & 1.53    & 3.79       \\ 
   \hline
  \end{tabular}

  \begin{tabular}{||c|c|l|l|l|l|l|l||}
   \hline
   FE    & $V$    & 
   $D_{2-1}$ & $D_{4-2}$ & $D_{4-3}$ & 
   $D_{6-3}$ & $D_{6-5}$ & $D_{8-4}$ \\ \hline
   \multirow{3}{*}{ $u_e$ } 
    & $V_A$  & 1.55 & 2.53 & 4.26 & 3.84 & 4.73 & 4.84 \\ 
    & $V_B$  & 1.52 & 2.60 & 3.68 & 3.42 & 6.00 & 4.91 \\
    & $V_C$  & 1.59 & 2.59 & 3.98 & 3.85 & 6.34 & 5.61 \\ 
   \hline
   \multirow{3}{*}{ $u_1$ } 
    & $V_A$  & 1.73 & 1.95 & 1.94 & 1.95 & 1.95 & 1.95 \\ 
    & $V_B$  & 1.82 & 1.97 & 1.97 & 1.97 & 1.97 & 1.97 \\
    & $V_C$  & 1.47 & 1.47 & 1.47 & 1.47 & 1.47 & 1.47 \\ 
   \hline
   \multirow{3}{*}{ $u_2$ } 
    & $V_A$  & 1.55 & 2.52 & 3.81 & 3.87 & 2.85 & 4.86 \\
    & $V_B$  & 1.52 & 2.61 & 2.68 & 3.39 & 2.45 & 3.35 \\
    & $V_C$  & 1.60 & 3.15 & 3.91 & 3.93 & 3.82 & 4.06 \\ 
   \hline
   \hline
   FE    & $V$    & 
   $DD_{2-1}$ & $DD_{4-2}$ & $DD_{4-3}$ & 
   $DD_{6-3}$ & $DD_{6-5}$ & $DD_{8-4}$ \\ \hline
   \multirow{3}{*}{ $u_e$ } 
    & $V_A$  & 0.55 & 1.52 & 3.12 & 2.54 & 4.34 & 3.81 \\ 
    & $V_B$  & 0.51 & 1.60 & 2.65 & 2.36 & 4.69 & 3.72 \\ 
    & $V_C$  & 0.50 & 1.54 & 3.52 & 2.82 & 5.29 & 4.34 \\ 
   \hline
   \multirow{3}{*}{ $u_1$ } 
    & $V_A$  & 0.51 & 0.99 & 0.95 & 0.96 & 0.90 & 0.96 \\ 
    & $V_B$  & 0.49 & 1.26 & 0.93 & 1.01 & 0.87 & 0.97 \\ 
    & $V_C$  & 1.10 & 1.00 & 0.38 & 1.04 & 0.47 & 0.21 \\ 
   \hline
   \multirow{3}{*}{ $u_2$ } 
    & $V_A$  & 0.55 & 1.51 & 2.89 & 2.56 & 2.06 & 3.82 \\ 
    & $V_B$  & 0.52 & 1.60 & 2.06 & 2.33 & 1.61 & 2.86 \\ 
    & $V_C$  & 0.50 & 1.54 & 2.60 & 2.83 & 3.09 & 3.43 \\ 
   \hline
  \end{tabular}
  } % end small
  \end{center}
  \caption{Top table: convergence orders of the solution error in the $l_2$-norm, for
           linear and quadratic finite elements, for the three test potentials.
           Middle and bottom tables: convergence orders of the first and second
           numerical derivatives, computed with different SBP operators.}
  \label{T:ConvNDXL2}
  \end{table}
% +----------------+----------------+----------------+------------------+----------------+----------------+

% +----------------+----------------+----------------+------------------+----------------+
  \begin{table}
  \begin{center}
  {\small
  \begin{tabular}{||c|c|c||}
   \hline
   potential   & linears & quadratics \\ \hline
   $V_A$       & 1.92    & 3.68       \\ \hline
   $V_B$       & 1.98    & 3.37       \\ \hline
   $V_C$       & 1.52    & 3.76       \\ 
   \hline
  \end{tabular}

  \begin{tabular}{||c|c|l|l|l|l|l|l||}
   \hline
   FE    & $V$    & 
   $D_{2-1}$ & $D_{4-2}$ & $D_{4-3}$ & 
   $D_{6-3}$ & $D_{6-5}$ & $D_{8-4}$ \\ \hline
   \multirow{3}{*}{ $u_e$ } 
    & $V_A$  & 0.99 & 2.02 & 4.70 & 3.33 & 4.32 & 4.69 \\ 
    & $V_B$  & 0.90 & 1.99 & 3.15 & 2.92 & 5.37 & 3.98 \\ 
    & $V_C$  & 1.04 & 2.14 & 3.63 & 3.47 & 6.26 & 4.99 \\ 
   \hline
   \multirow{3}{*}{ $u_1$ } 
    & $V_A$  & 0.95 & 0.99 & 0.98 & 0.98 & 0.98 & 0.98 \\ 
    & $V_B$  & 0.97 & 0.65 & 0.84 & 0.82 & 0.83 & 0.84 \\ 
    & $V_C$  & 1.47 & 1.47 & 1.47 & 1.47 & 0.55 & 1.46 \\ 
   \hline
   \multirow{3}{*}{ $u_2$ } 
    & $V_A$  & 1.00 & 2.02 & 2.13 & 3.19 & 2.00 & 4.52 \\ 
    & $V_B$  & 0.91 & 1.99 & 2.09 & 2.73 & 1.96 & 2.40 \\ 
    & $V_C$  & 1.04 & 2.09 & 3.03 & 3.70 & 3.13 & 3.35 \\ 
   \hline
   \hline
   FE    & $V$    & 
   $DD_{2-1}$ & $DD_{4-2}$ & $DD_{4-3}$ & 
   $DD_{6-3}$ & $DD_{6-5}$ & $DD_{8-4}$ \\ \hline
   \multirow{3}{*}{ $u_e$ } 
    & $V_A$  &  0.09 & 1.02 & 2.27 & 1.80 & 4.31 & 3.77 \\
    & $V_B$  & -0.07 & 0.98 & 2.16 & 1.95 & 4.21 & 2.79 \\
    & $V_C$  &  0.05 & 1.10 & 2.09 & 2.29 & 4.89 & 3.80 \\
   \hline
   \multirow{3}{*}{ $u_1$ } 
    & $V_A$  &  0.11 &-0.06 &-0.05 &-0.04 &-0.07 &-0.04 \\
    & $V_B$  & -0.07 &-0.27 &-0.17 &-0.10 &-0.22 &-0.08 \\
    & $V_C$  & -0.47 &-0.55 &-0.60 &-0.54 &-0.61 &-0.92 \\
   \hline
   \multirow{3}{*}{ $u_2$ } 
    & $V_A$  &  0.06 & 1.08 & 1.07 & 1.33 & 0.92 & 1.13 \\
    & $V_B$  & -0.07 & 0.99 & 1.26 & 1.55 & 0.92 & 1.63 \\
    & $V_C$  &  0.05 & 1.11 & 2.42 & 1.65 & 2.56 & 2.41 \\ 
   \hline
  \end{tabular}
  } % end small
  \end{center}
  \caption{This table displays the same type of information of the previous one, but this
    time in the $l_{\infty}$ norm. }
  \label{T:ConvNDXL8}
  \end{table}
% +----------------+----------------+----------------+------------------+----------------+

The results of our numerical experiments are illustrated by the
figure~\ref{F:cos3-ndx-norm2} and summarized in tables~\ref{T:ConvNDXL2}
and~\ref{T:ConvNDXL8}. These tables list convergence orders in the $l_2$ and
$l_{\infty}$ norms for each SBP operator, applied once and twice to three
different grid functions: the exact solution, the numerical solution obtained
with linear elements, and the numerical solution obtained with quadratics.
Test problem is the equation~\ref{Eq:BrillWave1} with three test potentials:
$V_A$ with $\omega=0.1$, $V_B$ with $r_0=4$ and $V_C$ with parameters
$\sigma_r=10$, $\sigma_z=5$, $r_0=1.5$. For the potentials $V_A$ and $V_B$ we
use a spherical domain with 7 patches, $R_{out}=10$, $a_c=2.5$, and grid size
ratio $N~:~N_r=1~:~2$.  For $V_C$ we use 13 patches with $R_{out}=20$,
$R_{med}=7$, $a_c=1.5$, and grid size ratios
$N~:~N_{r,in}~:~N_{r,out}=1~:~1~:~1$.

Figure~\ref{F:cos3-ndx-norm2} shows in more details some of this information, 
displaying log-log plots of the $l_2$-norms of
the errors of the solution and its first and second numerical derivatives, computed
with various FD operators, for the problem~\ref{Eq:BrillWave1} with the
potential $V_A$.

Several conclusions can be drawn from these tables and figure. In general, one
expects the convergence order of the first FD derivative to be the smallest
between the convergence order of the finite element solution itself and the
convergence order of the SBP operator used to compute the derivative. Our
results support this expectation: a) the first numerical derivative of the
\emph{exact solution} converges with the order of SBP operator used to compute
it. b) The  convergence order for the first derivative of the numerical
solution obtained with linear elements approaches $2$ for all SBP operators.
c) The convergence order for the first derivative using quadratics  improves
as the SBP order increases. Eventually, when the operators $D_{8-4}$ and
$D_{6-5}$ are used, the convergence order is either equal to the convergence
of the finite element solution, or to the convergence order of the
corresponding SBP operator.

The numerical results also show that second numerical differentiation takes
away one order of convergence for all three grid functions. In particular, the
second numerical derivative of linear elements solution fails to converge in
the $l_{\infty}$ norm (see table~\ref{T:ConvNDXL8}). Similarly, the second
numerical derivative computed with the $D_{2-1}$ operator fails to converge as
well.

The most important conclusion from these results is that taking successive
numerical derivatives of the grid function is only efficient for quadratics,
i.e. when superconvergence takes place. In this case, the finite element error
can be smaller than finite differencing one, and the convergence rate of the
latter is observed. For linear elements (and other elements of odd order),
other techniques are required. One of the possible solutions is
superconvergent gradient recovery, when the finite element solution is
differentiated and then its discontinuous derivative is projected back onto
original finite element space (sometimes with additional postprocessing,
see, for example, 
\cite{Levine85:superconv-recovery-grad},
\cite{Goodsel89:unified-superconv-grad},
\cite{KvrivzekNeittaanmaki87:global},
\cite{Blum86:richardson-grad-recov}
)
We do not pursue this direction, because superconvergent
gradient recovery for linear element solution will only produce no more than
second-order initial data, while for quadratics we already have third-order
convergent initial data without extra effort.

%-----------------------------------------------------------------------------
\subsection{Adaptive mesh refinement}
\label{SS:AMR}
\begin{figure}[htbp]
   \begin{center}
    \begin{tabular}{cc}
    \includegraphics[width=0.55\textwidth]{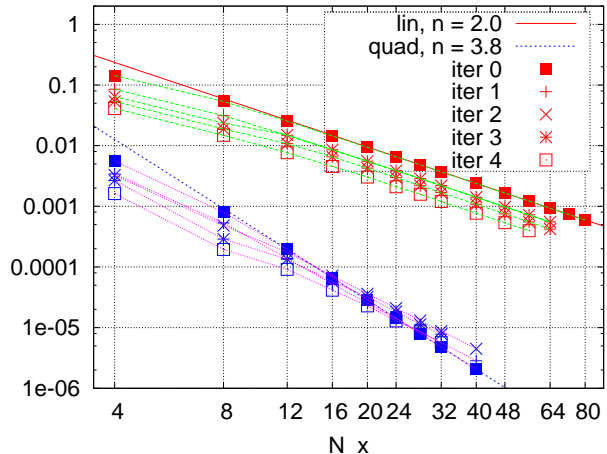}
    \end{tabular}
    \caption{
      Error in the $l_2$-norm for the solutions obtained with linear and and quadratic finite
      elements with adaptive mesh refinement. The test problem used is that
      one of section~\ref{SS:SecOrdConvSol} with potential $V_C$.
      The straight lines represent the fitted convergence exponents.
      For linear elements, at each resolution we started with a 
      semi-structured multi-block triangulation and did four AMR
      iterations. The $l_2$ norms of errors for these interations are
      plotted with the same abscissa, corresponding to the initial
      resolution. For each resolution, the set of iterations done with
      quadratic finite elements is the result of $p$-refinement on the
      linear element meshes.
    }
    \label{F:err-amr-tr}
   \end{center}
  \end{figure}

%-----------------------------------------------------------------------------
One of the main advantages of the finite element method in general and~\FETK{}
in particular is fully adaptive mesh refinement (AMR). We have explored AMR in
our semi-structured grids through the following strategy: we start from a
multi-block triangulation, adaptively refine it until the error reaches some
predefined level, and read off the final result at the original multi-block
triangulation nodes ~\ref{F:err-amr-tr}. In doing so, we have learned that for
the type of problems we are solving for in this paper, AMR is not necessarily
the most efficient strategy.

% /---------------------------------------------------------------------\
% \---------------------------------------------------------------------/

First of all, no matter how much the mesh is refined, in the end the finite
element solution has to be restricted to the same FD grid. What matters for
our purposes is the ability of the FD code to properly approximate the
solution on this grid, and not at the refined FE mesh. The original FD
multi-block grid is already adapted to be sufficiently fine to resolve all
important features of the solution.

Second, in the cases here treated both the final solution  is a rather
regular smooth function, and the domain has regular boundaries. For such cases,
standard uniform refinement gives just as good error reduction as AMR.
AMR should be advantageous, though, in cases where, for example, the solution
has non-smooth features or the domain has non-smooth boundaries.

Finally, the set of grid points where we sample the solution (that is, the
multi-block ones) is very special. As already explained, the convergence
order at these points is generally higher than one might expect. The special
status of these grid points implies local symmetry of the finite element
function spaces with respect to these points, which in turn implies
superconvergence. AMR can easily break this symmetry and degrade both the
error and the convergence order to the level of ``ordinary'' points. If we
wanted to keep this symmetry and have superconvergence, we would have to
refine at least within the entire patch, but this would be almost as expensive
as refining the entire domain uniformly.

Our experiments do not show particular advantage of using AMR compared to the
uniform refinement of semi-structured multi-block triangulations. In our
experiments we found that, in general, after a few refinement steps the AMR
error saturates, while uniform refinement error continues to decrease. We have
found that often, while decreasing the global $l_2$ norm, AMR leads to an
increase in the $l_{\infty}$ norm of the solution, because the local symmetry
is broken at several grid points in the refined mesh.

We have also tried a simple form of $p$-refinement.  Namely, using the same
set of refined meshes for linear and quadratic elements (see
figure~\ref{F:err-amr-tr}). We chose several initial multi-block triangulation
meshes with different resolutions and did four AMR iterations with linear
elements. Then we used the same meshes to increase the order of finite
elements to quadratics ($p$-refinement). It turns out that at high resolutions
the meshes which give the best error reduction when going from linear to
quadratic elements are initial triangulation meshes with no AMR. It happens so 
because the refined mesh loses the property of local symmetry at some of the grid
points, the pointwise superconvergence for quadratic elements is lost and the
global $L_2$-norm error is observed instead.

%=============================================================================
\section{Brill waves initial data and evolutions}
\label{S:BrillWaveID}
%-----------------------------------------------------------------------------

In General Relativity, initial data on a spatial 3D-slice has to satisfy the
Hamiltonian and momentum constraint equations~\cite{Arnowitt62,MisnerHoopRef},
\begin{align}
{}^3R - K^{ij}K_{ij} + K^2 &= 0\\
\nabla_i(K^{ij} - g^{ij}K) &= 0
\end{align}
where $K_{ij}$ and $K$ are the \textit{extrinsic curvature} of the 3D-slice and
its trace, respectively, and ${}^3R$ the Ricci scalar associated with the
spatial metric $g_{ij}$.

Brill waves~\cite{Bri59} 
constitute a simple yet rich example of initial data in numerical
relativity. In such a case  the extrinsic curvature of the slice is zero, and
the above equations reduce to a single one, stating that the Ricci scalar has to
vanish:
\begin{align}
 {}^3R = 0 \,. \label{Eq:Ris0}
\end{align}

If the spatial metric is given up to one unknown function, Eq.~(\ref{Eq:Ris0})
in principle allows us to solve for such function and thus complete the construction 
of the initial data. The Brill equation is a special case of~(\ref{Eq:Ris0}), where 
the 3-metric is expressed through the conformal transformation $g_{ij} = \psi^4
\tilde{g}_{ij}$ of an unphysical metric $\tilde{g}_{ij}$, with an unknown
conformal factor $\psi$. Equation~\eqref{Eq:Ris0} then becomes~\cite{B:Mur93}:
\begin{align}
 (-\nabla_{\tilde{g}}^2 + \frac{1}{8}\tilde{R})\psi = 0
  \label{Eq:BrillWave2}
\end{align}
where $\tilde{R}$ and $\nabla_{\tilde{g}}^2$ are the Ricci scalar curvature
and Laplacian of the unphysical metric $\tilde{g}_{ij}$, respectively.

Here we will focus on the axisymmetric case with the unphysical metric given in
cylindrical coordinates by

\begin{align}
 \tilde{g}_{ij} = e^{2q(\rho,z)}(d\rho^2+dz^2) + \rho^2 d\ji^2 \,,
 \label{Eq:UnphMetric}
\end{align}
where $q(\rho, z)$ is a function satisfying the following conditions:
\begin{enumerate}
\item regularity at the axis: $q(\rho=0,z) = 0$, $\frac{\pd q}{\pd\rho}|_{\rho=0} = 0$,
\item asymptotic flatness: $q(\rho,z)|_{r\to\infty} < O(1/r^2)$, 
      where $r$ is the spherical radius $r~=~\sqrt{\rho^2+z^2}$ \,.
\label{L:AsymptoticFlatnessBC}
\end{enumerate}

The Hamiltonian constraint equation~(\ref{Eq:Ris0}) becomes a second order
elliptic PDE, which with asymptotically
flat boundary conditions at $r\to\infty$ takes the form
\begin{eqnarray}
\label{Eq:BrillWave}
-\nabla^2\psi(\rho,z) + V(\rho,z)\psi(\rho,z) & = & 0, \\
\label{Eq:PsiBoundaryCond}
 \psi|_{r\to\infty} &=& 1 + \frac{M}{2r} + O(1/r^2),
\end{eqnarray}
with the potential $V(\rho,z)$ given by 
$$
 V = -\frac{1}{4}(q''_{\rho\rho} + q''_{zz}).
$$

We numerically solve this equation using~\FETK{} on the 13-patch multi-block spherical domain described
 in section~\ref{SS:DomStr} (see figure~\ref{F:patchsystems}).  We use domain 
 parameters $R_{out}~=~30$, $R_{med}~=~7$, $a_c~=~1.5$, and grid dimension
ratios $N:N_{r,inner}:N_{r,outer}=2:3:12$. Our low-medium-high resolution
triple is $N=32$, $N=36$ and $N=40$, except for pointwise convergence tests on
the $x$-axis (see figure~\ref{F:pointwise-conv-cpsi}), where we use $N=16$,
$N=24$ and $N=36$ (since they all differ by powers of $1.5$).

We impose Robin boundary conditions, as in equation~\eqref{Eq:RobinBC}. The
weak form~\eqref{Eq:WeakFormRobinBC} and bilinear linearization
form~\eqref{Eq:BilinearLinearizationRobinBC} for this problem are given in
section~\ref{SS:SecOrd}. Since first order elements lead to unacceptably low
convergence orders for most general relativistic applications, from hereon we
restrict ourselves to quadratic ones (which should give fourth order
convergence if superconvergence is exploited).

% /---------------------------------------------------------------------\
  \begin{figure}[htbp]
  \begin{center}
  \begin{tabular}{cc}
  \includegraphics[width=0.45\textwidth]{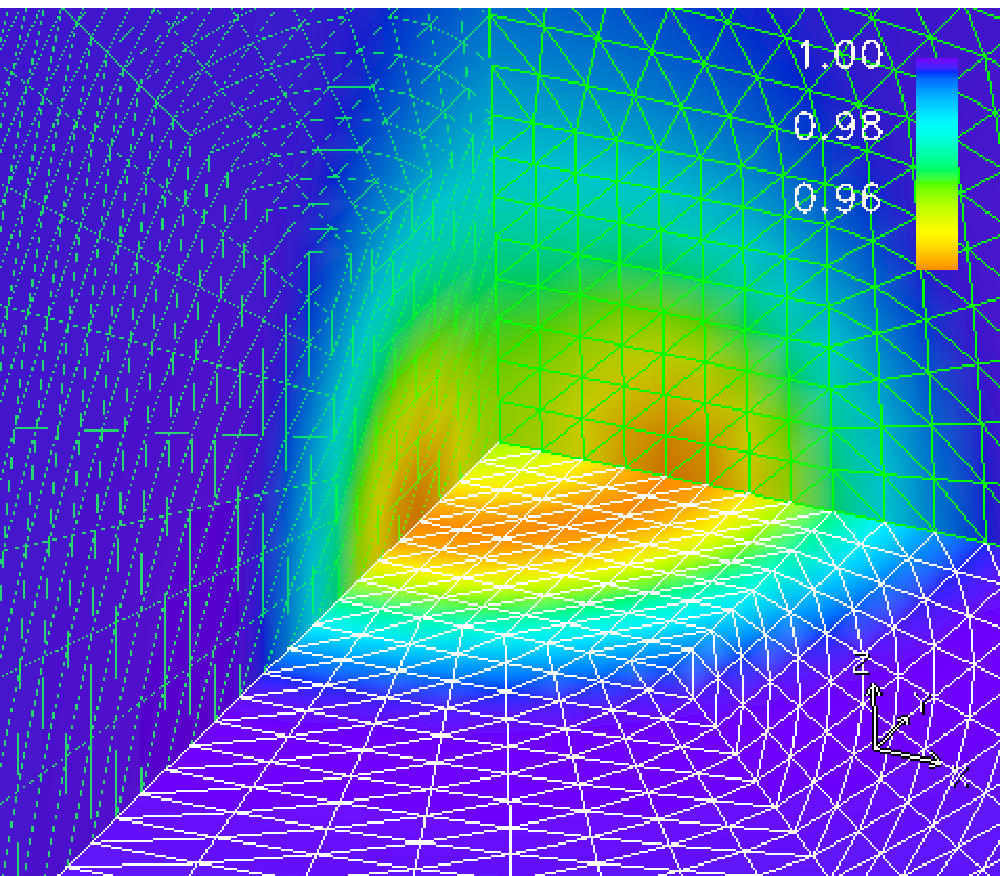}     &
  \includegraphics[width=0.45\textwidth]{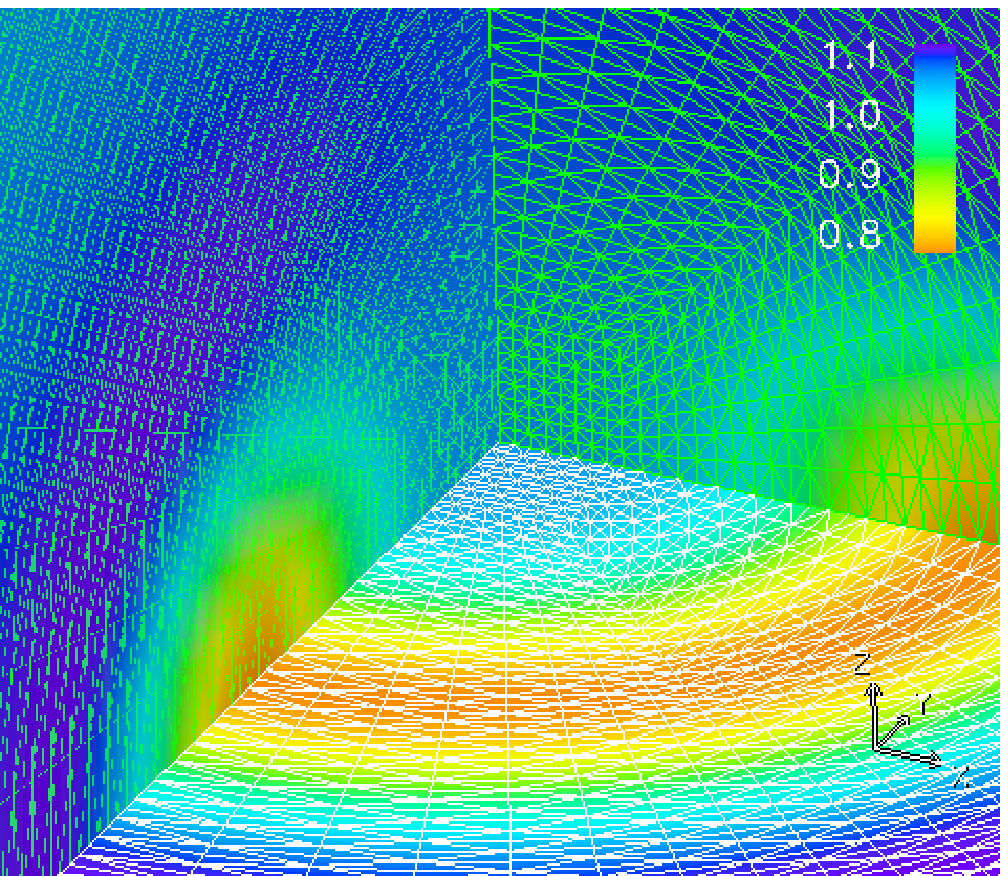} \\
  {\tiny (a)} & {\tiny (b)}
  \end{tabular}
  \caption{Potentials for the two types of Brill waves considered: 
           Holz' (a) and  toroidal (b) forms.}
  \label{F:bw-examples}
  \end{center}
  \end{figure}
% \---------------------------------------------------------------------/

We work with two specific choices for
$q(\rho,z)$:
\begin{itemize}
\item[(a)] Holz' form~\cite{alcubierre-2000-61}: 
           $q_H(\rho,z) = a_H \rho^2 e^{-r^2}$, with amplitude $a_H = 0.5$;
\item[(b)] toroidal form:  
           $q_t(\rho,z) = a_t \rho^2 \exp{\left(
               -\frac{(\rho-\rho_0)^2}{\sigma_{\rho}^2}
               -\frac{z^2}{\sigma_z^2}
           \right)}$, 
           with amplitude $a_t = 0.05$, radius $\rho_0=5$, width in
           $\rho$-direction $\sigma_{\rho}=3.0$ and width in
           $z$-direction $\sigma_z=2.5$.
\end{itemize}

Holz potential is chosen for historical reasons. It is suitable for
evolutions using the cubed sphere system, because initially the wave is
concentrated near the origin (at the central patch), and later decays into a
sequence of spherical waves. The cubed sphere grid resolves the wave well both
at the initial moment, and during the subsequent evolution. For the toroidal
potential, the cubed sphere system is adapted even better, since it
efficiently removes the dependence on one of the angular coordinates, $\ji$.

%-----------------------------------------------------------------------------
\subsection{Importing the initial data into QUILT}
\label{SS:ImportID}
%-----------------------------------------------------------------------------
Once we have constructed the initial data sets, we analyze and evolve them by
importing them into QUILT. For evolutions we use the Generalized Harmonic (GH)
first order symmetric hyperbolic formulation of the Einstein's
equations introduced in~\cite{2006CQGra..23S.447L}, which features 
exponential suppression of short-wavelength constraint violations; our
multi-block implementation is described in \cite{Pazos:2006kz}.

The set of evolved variables in this system includes the 
4-metric $g_{ab}$, its spatial derivatives $\Phi_{iab}=\pd_ig_{ab}$ and
quantities $\Pi_{ab}=-t^c\pd_cg_{ab}$, where $t^c$ is the unit normal vector to the
spatial slice.

To set up the initial data, we first compute the $3+1$ quantities and then convert
them to GH variables. The 3-metric $g_{ij}(x,y,z)$ is computed from the conformal
factor $\psi(x,y,z)$, in Cartesian coordinates, using the expressions (which
follow from~\ref{Eq:UnphMetric}):
\begin{align}
  g_{xx} &= \psi^4(e^{2q}x^2+y^2)/\rho^2,  
 &g_{xy} &= \psi^4(e^{2q}-1)xy/\rho^2,    \notag \\
  g_{yy} &= \psi^4(x^2+e^{2q}y^2)/\rho^2,  
 &g_{zz} &= \psi^4 e^{2q},                \notag \\
  g_{xz} &= g_{yz} = 0 \notag
\end{align}

where $\rho=\sqrt{x^2+y^2}$.  Then we construct the rest of the evolved
variables, including the gauge source functions $H_a=-g^{bc}\Gamma_{a,bc}$
(here $\Gamma_{a,bc} = \frac{1}{2}(\pd_b g_{ac}+\pd_c g_{ab}-\pd_a g_{bc})$
are the Christoffel symbols). The extrinsic curvature is assumed to be zero,
$K_{ij}=0$, we also use unit lapse $\alpha=1$, zero shift $\beta_i=0$, and
zero time derivative of lapse and shift: $\pd_t\alpha=\pd_t\beta_i=0$.

%-----------------------------------------------------------------------------
\subsection{Convergence of initial data and Hamiltonian constraint}
\label{SS:ConvID}
\begin{figure}[htbp]
  \begin{center}
  \includegraphics[width=0.45\textwidth]{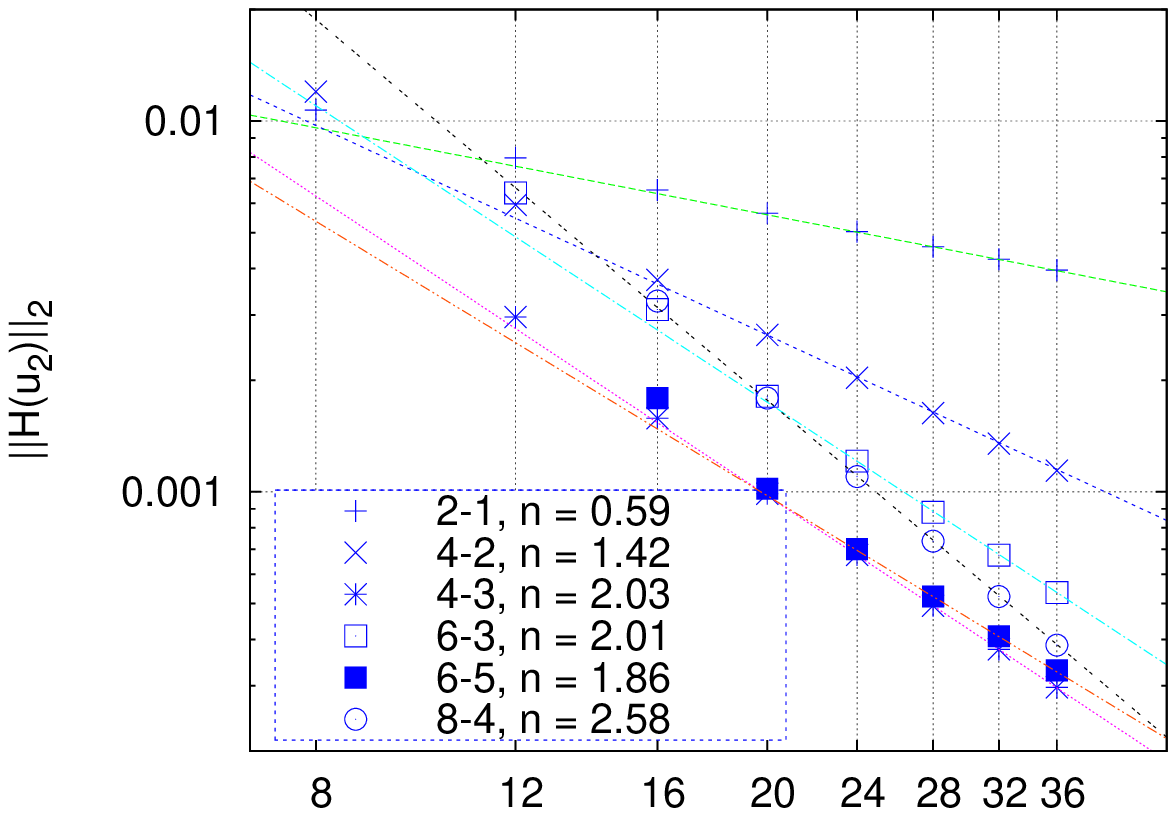} 
  \includegraphics[width=0.45\textwidth]{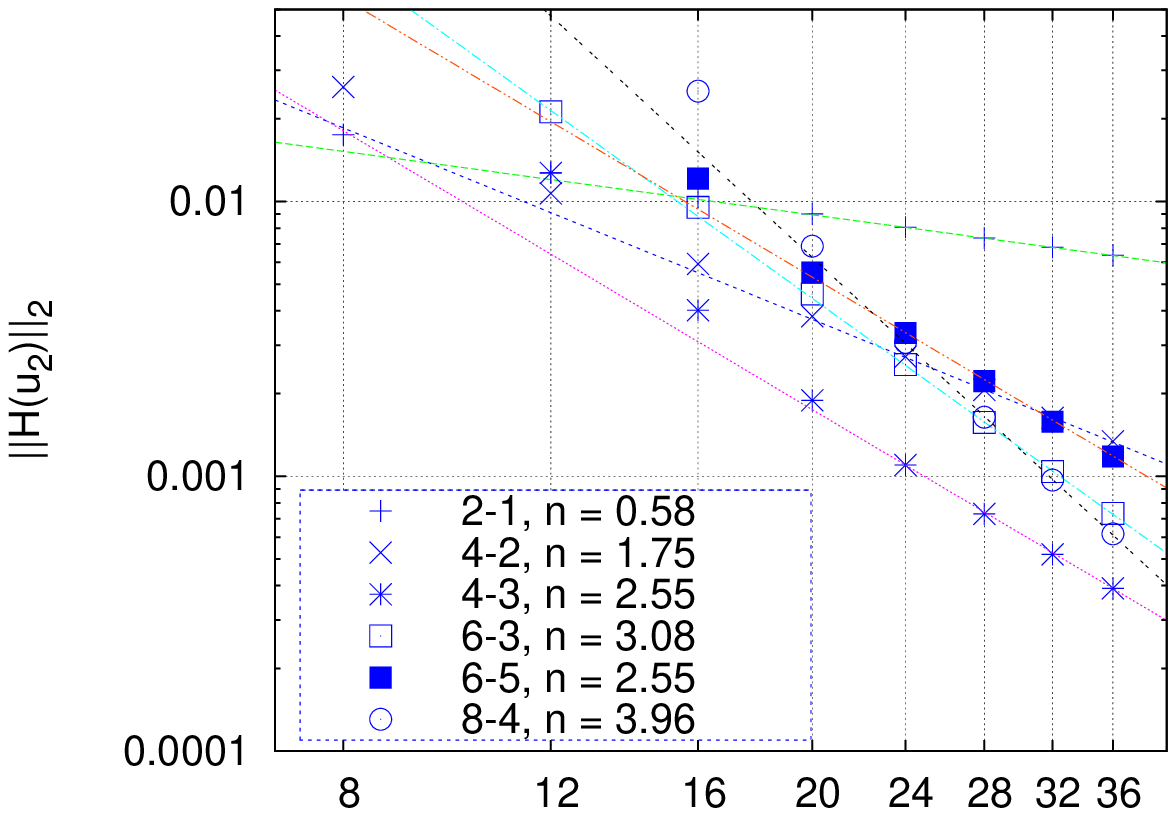} 
  \caption{
    Convergence of the Hamiltonian constraint in the $l_2$-norm for
    Brill waves of Holz (left) and toroidal (right) type, computed using
    quadratic finite elements and different numerical SBP operators.
  }
  \label{F:ham-nx-toroidal}
  \end{center}
  \end{figure}

%-----------------------------------------------------------------------------
To estimate the quality of our initial data, we evaluate the Hamiltonian constraint
violation using the SBP operators, which is in fact equivalent to an independent residual evaluation for
the Brill equation~\eqref{Eq:BrillWave}. As is the case with just numerical
derivatives, we find that the magnitude and convergence order of the Hamiltonian
constraint violation depends on both the order of the SBP operator, and the order
of finite elements. In total, computing the Hamiltonian constraint involves two
numerical differentiations, therefore it has to converge with the same order
as second numerical derivatives. We can see this convergence rate for different
SBP operators in figure~\ref{F:ham-nx-toroidal}. Table~\ref{T:ConvHam}
summarizes those results, along with the expected convergence orders
 for second numerical derivatives. 

Consistent with what one expects, 
for finite difference operators of sufficiently high order the constraints
 converge with the same order as the finite element solution
 itself, which should be $3+\sigma$ for some $0<\sigma\le1$ (depending on the level of
 superconvergence obtained). Similarly,  below in section~\ref{SS:ConvEvo} we will show that the 
 extracted gravitational waves have a similar convergence order.

% /---------------------------------------------------------------------\
% \---------------------------------------------------------------------/

% +----------------+----------------+----------------+----------------+----------------+----------------+----------------+
  \begin{table}
  \begin{center}
  {\small 
  \begin{tabular}{||c|l|l|l|l|l|l||}
   \hline
   $q$       &  
   $D_{2-1}$ & $D_{4-2}$ & $D_{4-3}$ & 
   $D_{6-3}$ & $D_{6-5}$ & $D_{8-4}$ \\ 
   \hline
    Holz     & 0.59 & 1.42 & 2.03 & 2.01 & 1.86 & 2.58 \\
   \hline
    Toroidal & 0.58 & 1.75 & 2.55 & 3.08 & 2.55 & 3.96 \\
   \hline
  \end{tabular} 
  }
  \end{center}
  \caption{Convergence orders of the Hamiltonian constraint for the two
           initial data sets in the $l_2$-norm.}
  \label{T:ConvHam}
  \end{table}
% +----------------+----------------+----------------+------------------+----------------+----------------+----------------+

% /---------------------------------------------------------------------\
  \begin{figure}[htbp]
  \begin{center}
  \begin{tabular}{cc}
  \includegraphics[width=0.45\textwidth]{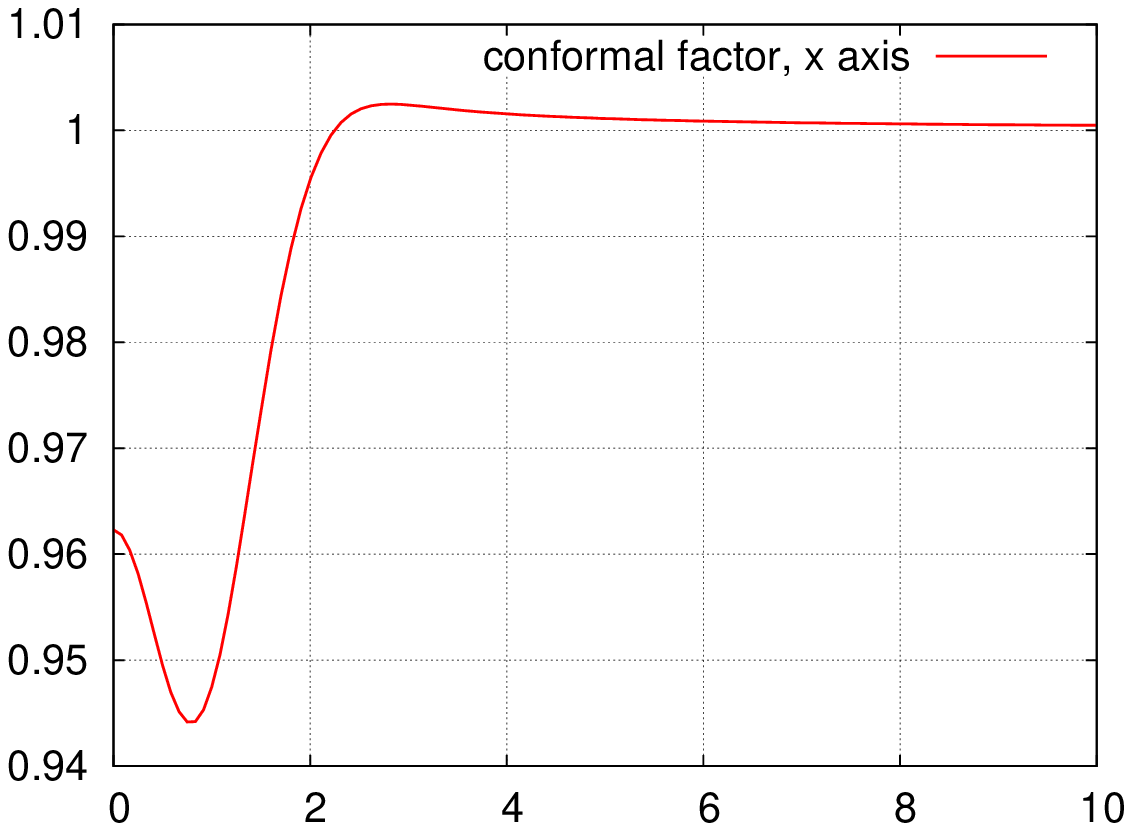} &
  \includegraphics[width=0.45\textwidth]{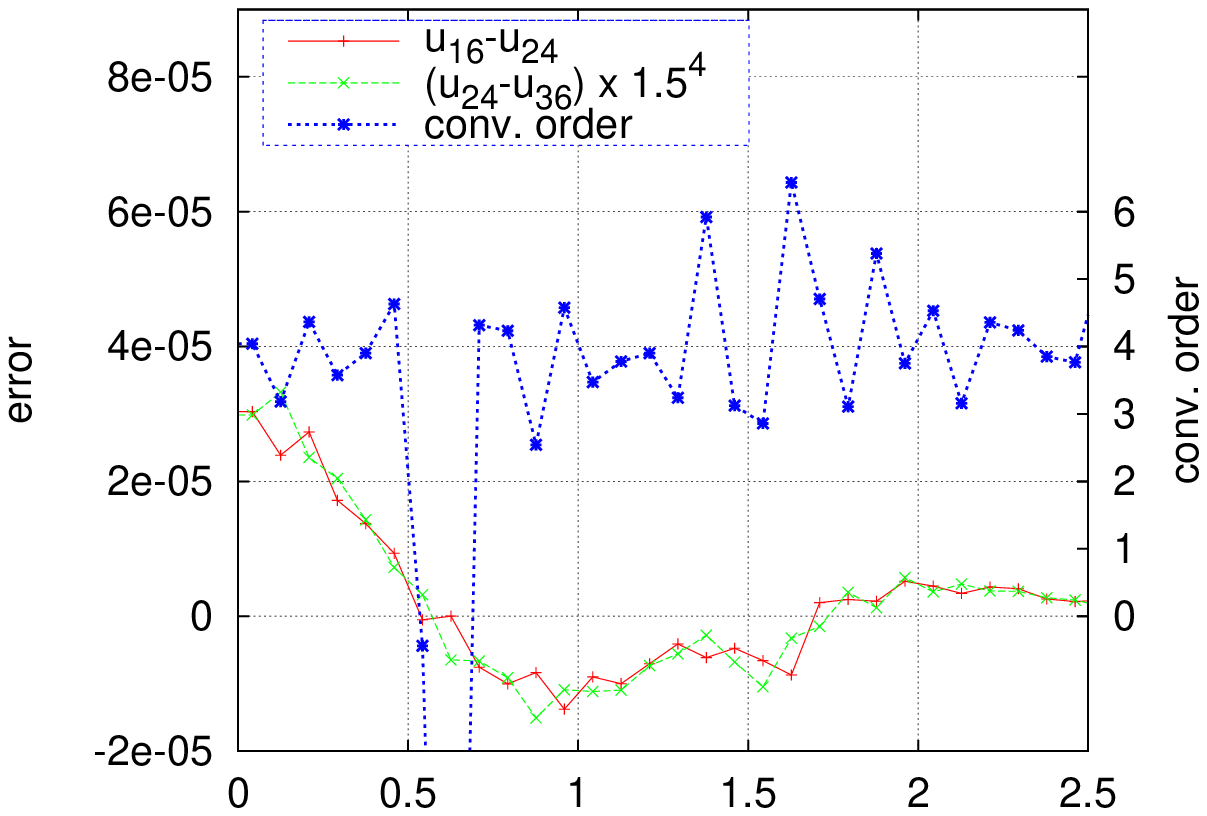} \\
  {\tiny (a)} & {\tiny (b)} \\
  \end{tabular}
  \caption{
    (a) 1-d cut through the Brill wave conformal factor $\psi_{fine}$
        for problem (a) along the $x$-axis.
    (b) Errors $\psi_{coarse}-\psi_{medium}$, 
        $\psi_{medium}-\psi_{fine}$, and pointwise convergence order on
        the 1-d cut along the $x$-axis.
  }
  \label{F:pointwise-conv-cpsi}
  \end{center}
  \end{figure}
% \---------------------------------------------------------------------/

It is not difficult to see why the convergence order is lower for problem
(a). This is the order we might expect for quadratic elements on completely
unstructured meshes, without superconvergence. The reason that no
superconvergence is observed in this case is the following. We have chosen the
domain and the width of the Brill wave in such a way that around the boundary
of the inner cubical patch the solution varies significantly (see
figure~\ref{F:pointwise-conv-cpsi}). Recall that the
size of that patch is $a_c=1.5$ and the width of the gaussian in the function
$q$ is $1$. But this is exactly the place where the local symmetry property is
violated most (especially at the corners of the cube), and conditions for
superconvergence are the least favorable. Everywhere else the solution varies
very slowly and the error is small compared to the error at the boundary of
the central cubical patch. With increasing resolution this error 
dominates in both $l_2$ and $l_{\infty}$ norms.
The situation is better for the problem with toroidal potential, because most
of the variation of the solution is located outside the central cube (recall
that the radius of the toroidal wave we use is $r_0 = 5$); though this is still in the
inner-patches region (the radius of the spherical boundary between inner and
outer patches is $r_{med} = 7$). 

%-----------------------------------------------------------------------------
\subsection{Multi-block evolutions}
\label{SS:ConvEvo}
%-----------------------------------------------------------------------------
We now demonstrate that our approach for generating initial data on multi-block
grids using finite element methods can be successfully used in practice in 
fully nonlinear relativistic simulations. We do so by presenting results of
multi-block evolutions of the Holz set of Brill initial waves constructed above. 

In the notation of~\cite{2006CQGra..23S.447L}, we fix the 
damping parameters of the GH formulation to $\gamma_0=\gamma_2=1$. We used the SBP operator $D_{6-3}$ for
spatial differentiations, a 4-th order Runge-Kutta time integrator with adaptive
time stepping, and maximally dissipative outer boundary conditions.
 
\begin{figure}[htbp]
  \begin{center}
  \begin{tabular}{c}
  \includegraphics[height=0.3\textheight]{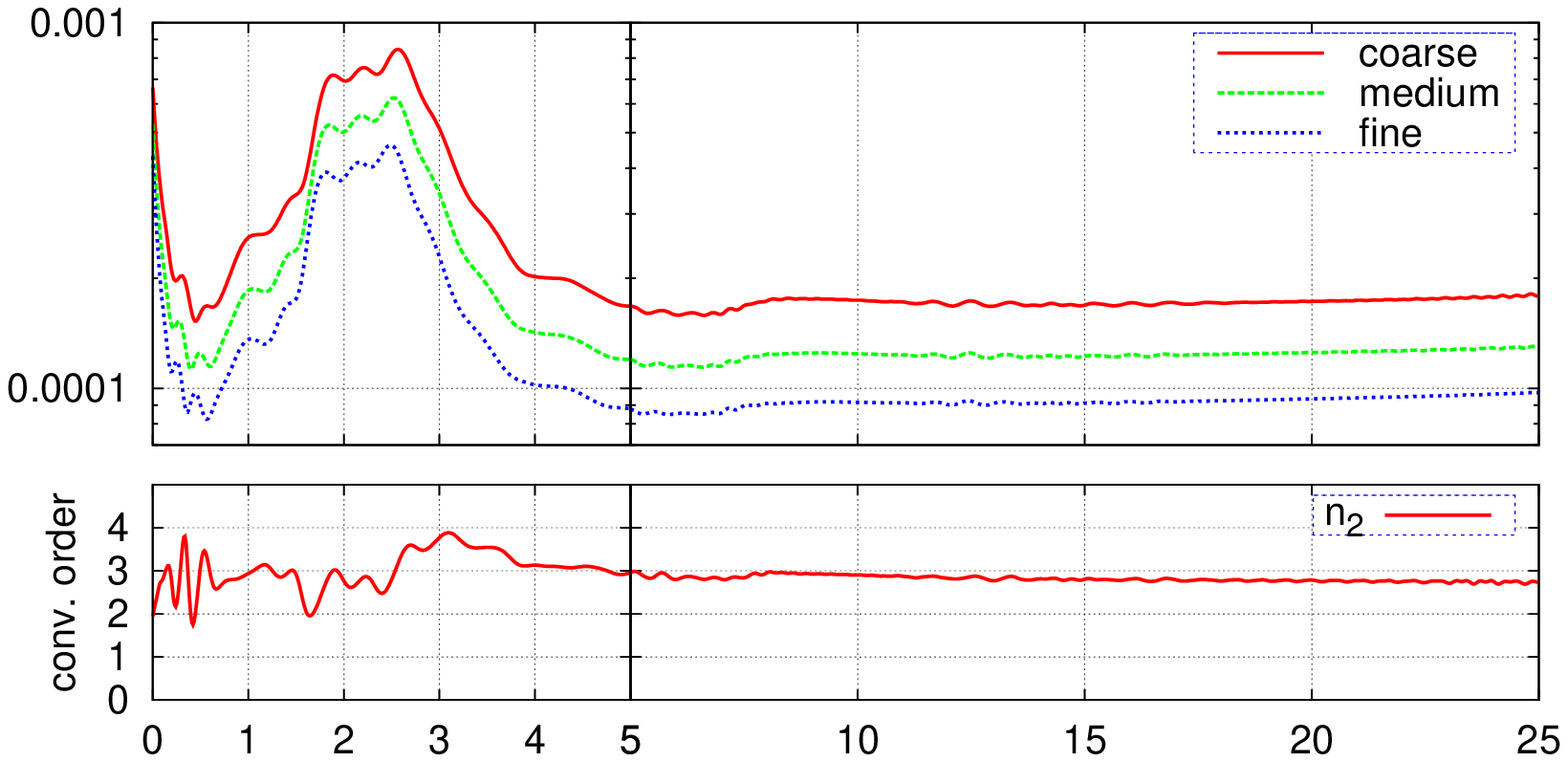}
  \end{tabular}
  \caption{
    Convergence of the Hamiltonian constraint as a function of time during a multi-block evolution
    of the Brill wave initial data generated with finite element methods. }
  \label{F:ham-conv-evo}
  \end{center}
  \end{figure}

The Brill wave amplitude $a_H=0.5$ is in the subcritical regime (the critical
value is around $a_{cr}\approx 4.85$~\cite{alcubierre-2000-61}). As a result 
the wave, initially concentrated near the center, dissipates and leaves the
domain after a while.  Figure~\ref{F:ham-conv-evo} shows a convergence plot in
time for the Hamiltonian constraint during such evolution. We see that the
Hamiltonian constraint converges with a factor of $2-3$ in the $l_2$ norm. 
This has to be one order less convergent
than the solution itself, therefore we anticipate the solution to
converge with a factor of $3-4$, which is in agreement with the
pointwise convergence order of the conformal factor at the initial time
(figure~\ref{F:pointwise-conv-cpsi}).

% /--------------------------------------------------------------------
% \---------------------------------------------------------------------/

We compute gravitational waveforms using a generalized Regge-Wheeler-Zerilli 
formalism, along the lines of ~\cite{Pazos:2006kz} and extending 
that reference to the even parity sector (details about that extension will be
presented elsewhere). The 
four-dimensional spacetime metric is decomposed into a spherically symmetric 
``background'' plus a small perturbation~\cite{Regge57,Zerilli70}. 
 The background part of the metric is identified with 
the Schwarzschild solution and the radiation content with the 
difference between the numerically computed solution and the background.

The  background metric is written as 
\begin{equation}
ds^2 = \tilde{g}_{\rho\tau}(t,r) dx^{\rho} dx^{\tau} + f^2(t,r) \hat{g}_{AB} dx^A dx^B.
\end{equation}
with the four-dimensional background manifold split as the
product space of a two-dimensional one $\Mfold$
endowed with coordinates $x^{\rho}$ (with $\rho,\tau=0,1$ usually denoting the time and
radial coordinate) and a
unit 2-sphere $S^2$ with coordinates $x^A$ (with $A=2,3$ commonly taken as
the $\theta$ and $\phi$ polar spherical coordinates).
Here $ \tilde{g}_{\rho\tau}$ is the metric of the manifold $\Mfold$ and
$f^2$ is a positive function of $x^{\rho}$. If using the areal radius as
a coordinate we have $f=r$. The metric of the 2-sphere is taken to
be $\hat{g}_{AB} = \textrm{diag}(1,\sin^2 \theta)$ in polar spherical coordinates.

The metric perturbation is decomposed in terms of scalar, vector and tensor spherical
harmonics~\cite{Zerilli70b,Thorne80b,Pazos:2006kz}.  The decomposition 
naturally splits the different $(\ell,m)$ modes into even $(-1)^\ell$ 
and odd $(-1)^{\ell+1}$ parity under reflections about the origin. The
two different parities are handled separately. Odd and even-parity perturbations
are described by the Regge-Wheeler~\cite{Regge57} and Zerilli~\cite{Zerilli70}
functions, respectively. 

\begin{figure}[htbp]
  \begin{center}
  \includegraphics[width=0.45\textwidth]{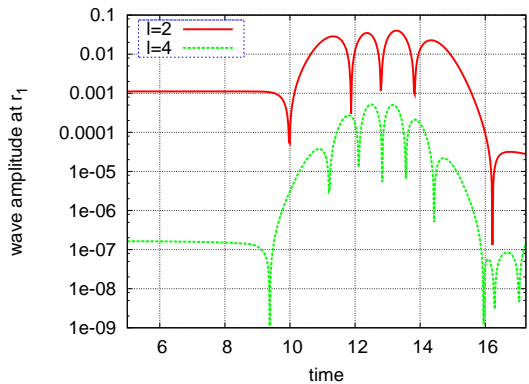} 
  \caption{
    Nonzero components of gravitational radiation $\ell=2$ and
    $\ell=4$, $m=0$, extracted at radius $r_{e}=12.75$.
  }
  \label{F:wave-all}
  \end{center}
  \end{figure}

\begin{figure}[htbp]
  \begin{center}
  \begin{tabular}{cc}
  \includegraphics[width=0.45\textwidth]{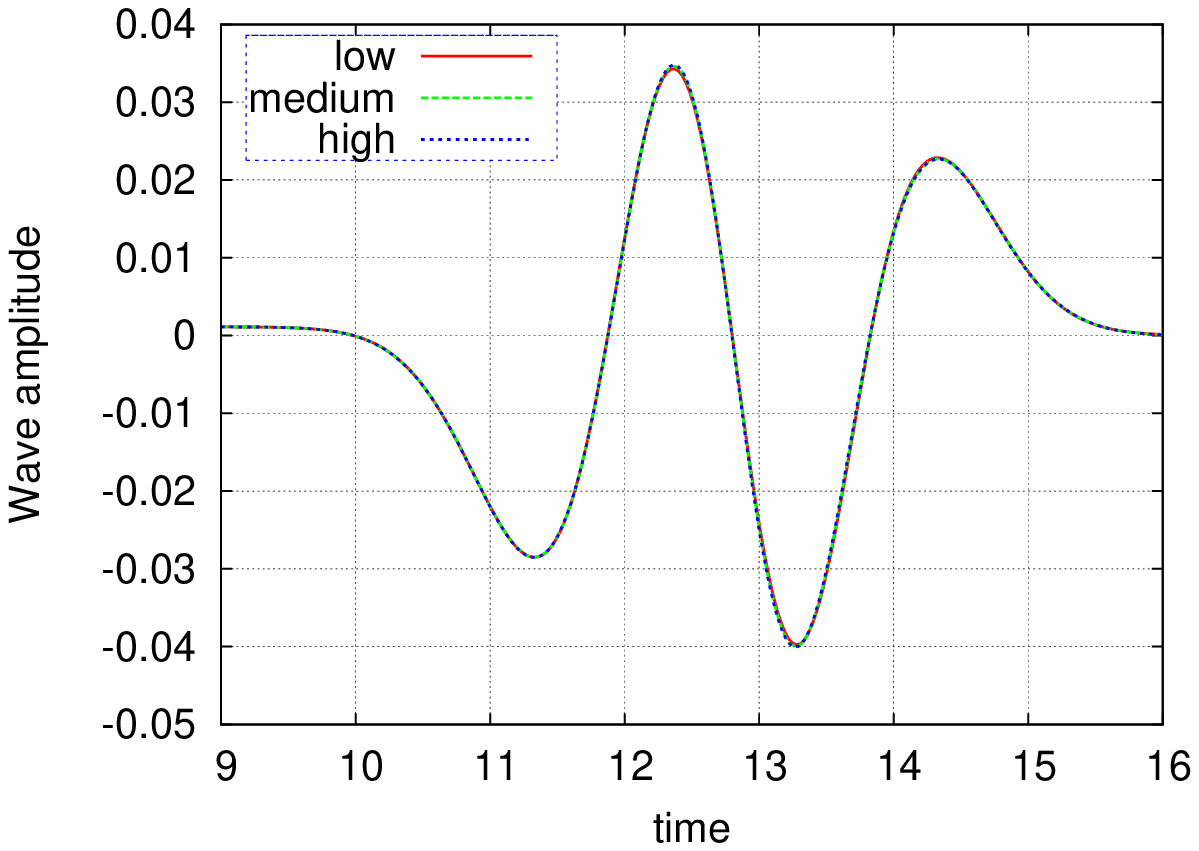} &
  \includegraphics[width=0.45\textwidth]{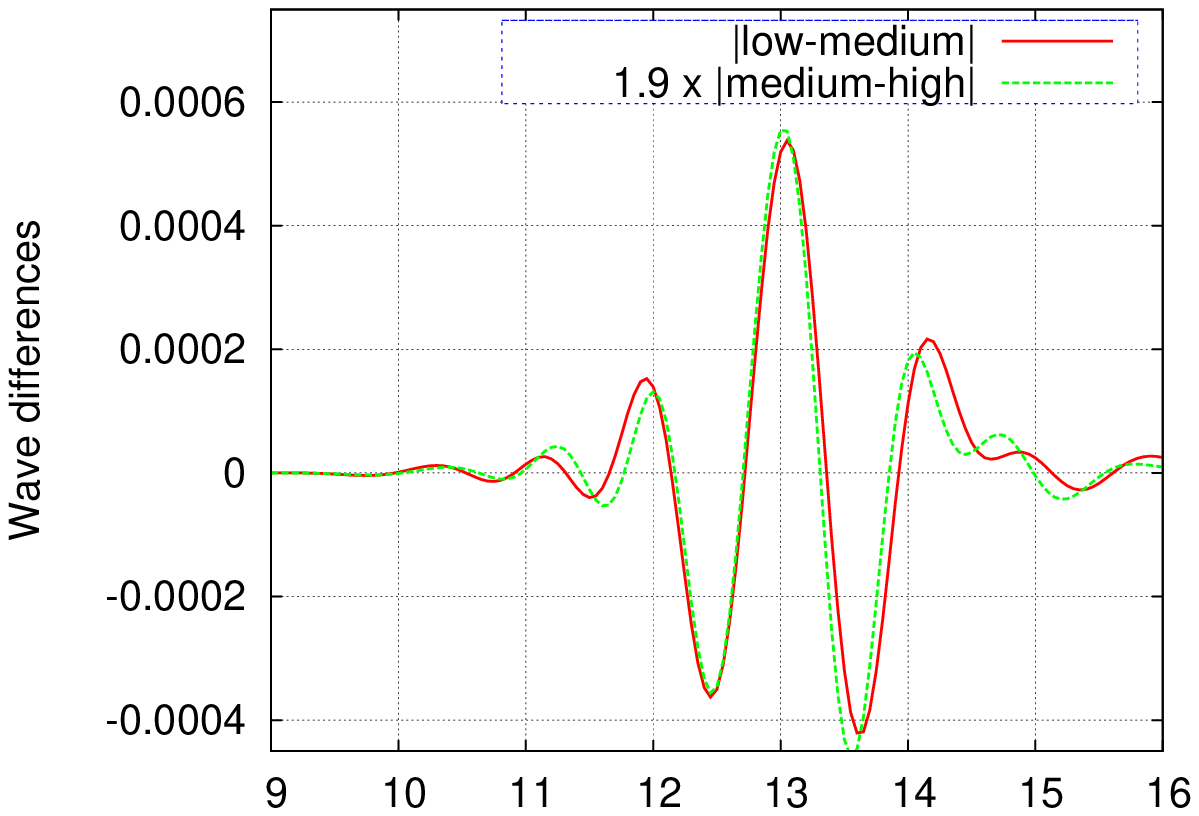} \\
  {\tiny (a)} & {\tiny (b)} \\
  \end{tabular}
  \caption{
 Zerilli function (left) and self-differences (right) for the
  $(\ell=2, m=0)$ mode, scaled according to fourth order convergence. 
  }
  \label{F:wave-l2}
  \end{center}
  \end{figure}

\begin{figure}[htbp]
  \begin{center}
  \begin{tabular}{cc}
  \includegraphics[width=0.45\textwidth]{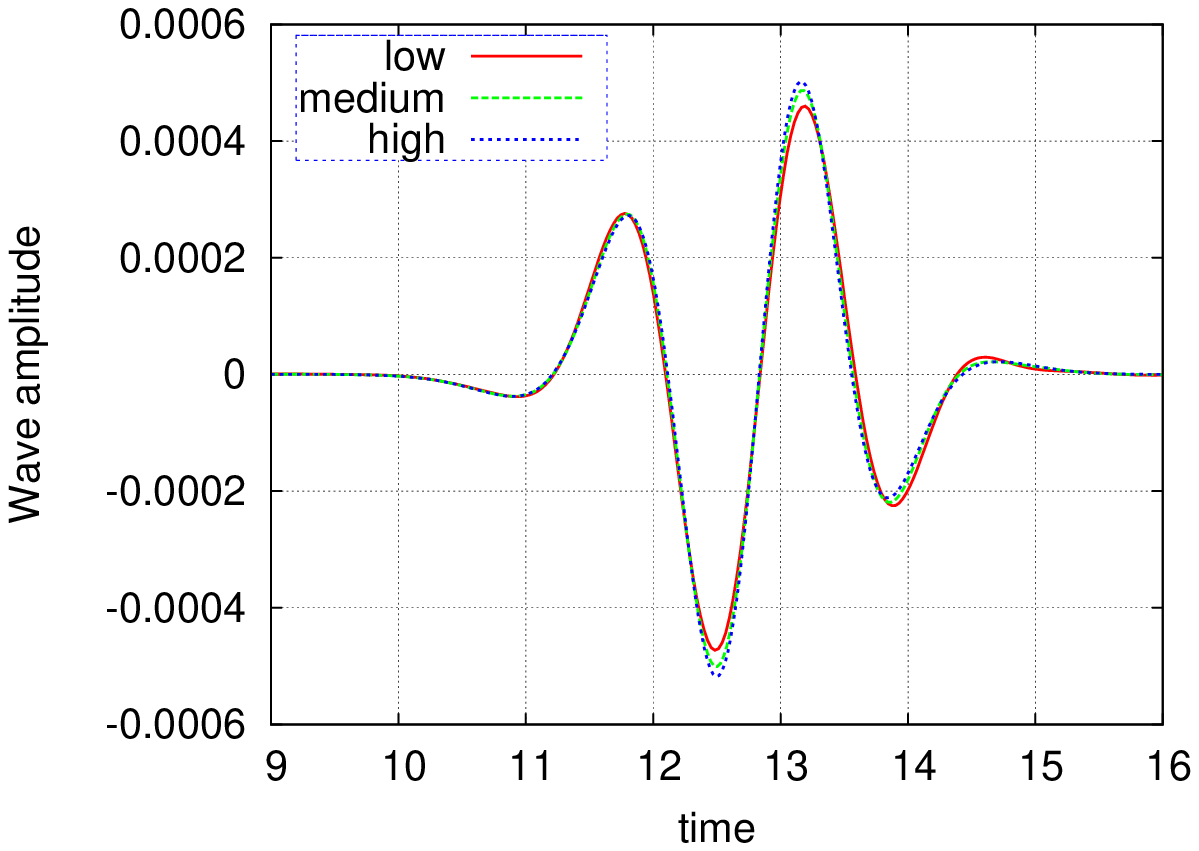} &
  \includegraphics[width=0.45\textwidth]{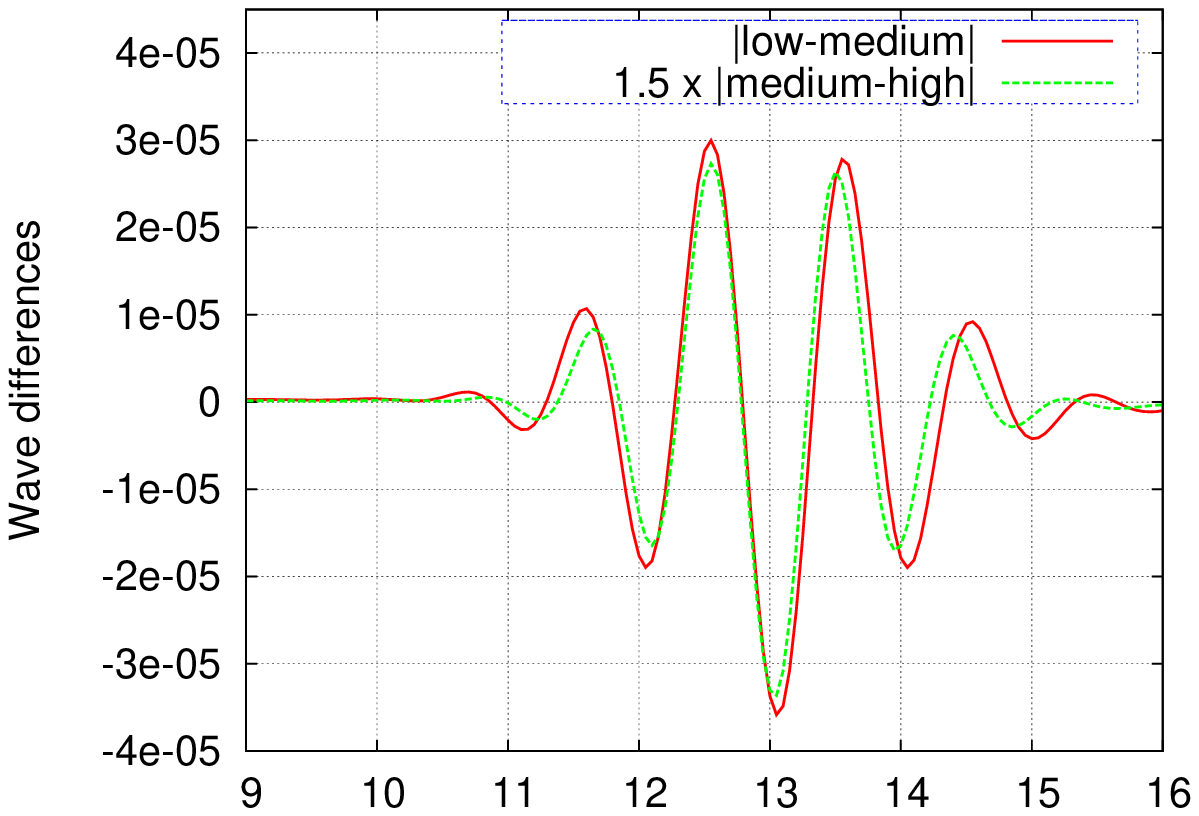} \\
  {\tiny (a)} & {\tiny (b)} \\
  \end{tabular}
  \caption{
    Zerilli function (left) and self-differences (right) for the
  $(\ell=4, m=0)$ mode, scaled according to third order convergence. 
  }
  \label{F:wave-l4}
  \end{center}
  \end{figure}

The dominant modes in the evolutions of the Brill data
constructed above are the $\ell=2, 4$ even parity,
axisymmetric ones (see figure~\ref{F:wave-all}). Figures \ref{F:wave-l2} and \ref{F:wave-l4} 
display the corresponding Zerilli functions and their convergence behavior,
extracted at a radius $r_e=12.75$.
 The observed convergence factors for the $\ell=2$ and $\ell=4$ modes
are around four and three, respectively, which are consistent with the
convergence factors from quadratic elements with superconvergence for the
initial data and the $D_{6-3}$ SBP operator and fourth-order Runge-Kutta for
the evolution.

% /---------------------------------------------------------------------\
% \---------------------------------------------------------------------/

% /---------------------------------------------------------------------\
% \---------------------------------------------------------------------/

% /---------------------------------------------------------------------\
% \---------------------------------------------------------------------/

%=============================================================================
\section*{Final remarks}
\label{S:Cnc}
%-----------------------------------------------------------------------------
In this paper we followed a finite element approach for generating initial
data satisfying the Einstein constraint equations on semi-structured,
multi-block three-dimensional domains.
In section~\ref{S:QualSol} we used semi-structured multi-block triangulations
to solve for some test problems with known closed-form solutions. The obtained
linear and quadratic finite element solutions were then restricted to the
multi-block grid, and their convergence, as well as the convergence of their
first and second derivatives, was evaluated numerically using independent
high-order finite difference operators satisfying summation by parts (SBP).
While the linear elements solution showed usual 2-nd order convergence
(unacceptably low for many relativistic applications), for quadratic elements
we obtained superconvergence with order $3+\sigma$ (with $0< \sigma\le1$) on
the grid, due to the approximate local symmetry at the mesh vertices with
respect to the local inversion of the multi-block triangulations.

Initial data for a first-order formulation of the Einstein equations
involves first derivatives of the solution of the constraints equation.
Computing the constraints or right-hand sides of the evolution equations
requires taking derivatives twice. In subsection~\ref{SS:SecOrdConvND} we
analyzed convergence of the first and second numerical derivatives, taken with
different high-order SBP operators. For quadratic elements, the first
numerical derivative was observed to converge with either the superconvergence
order $3+\sigma$, or the order of SBP operator. The latter is a transient
error behavior and happens when the FE error is still smaller than the error
of numerical differentiation.

In subsection~\ref{SS:AMR} we discussed three factors which make adaptive mesh
refinement (AMR) unnecessary and/or less efficient for the problems here
considered when compared to global refinement: the fact that the multi-block
grid is already tailored to resolve fine features of the solution, the need to
restrict the finite element solution to the same grid, and the
superconvergence properties of the quadratic elements solution. Because we
lose superconvergence when using completely unstructured meshes, adaptively
refining the solution sometimes makes the errors {\em larger} (see
figure~\ref{F:err-amr-tr} for an example).
However, we also noted that AMR would likely become advantageous for
other problems with more singular solution features.

Finally, in section~\ref{S:BrillWaveID} we presented numerical experiments
with Brill waves. The constraint equations in this case reduce to a single
elliptic one~\eqref{Eq:BrillWave} on the conformal factor $\psi$, which has to
be differentiated once to obtain the full set of initial data variables in the
generalized harmonic formulation (subsection~\ref{SS:ImportID}).
Subsection~\ref{SS:ConvID} presented a convergence analysis of the initial
data and Hamiltonian constraint, and confirmed that the initial data computed
with quadratic finite elements shows the desired order of convergence $> 3$
(see table~\ref{T:ConvHam}). Finally, in section~\ref{SS:ConvEvo} we
demonstrated stable, $>3$-rd order convergent multi-block evolutions of
subcritical Brill waves with finite differences, summation by parts operators,
and extracted the first two dominant radiation modes from the numerical
solution.

This paper shows that generating initial data on semi-structured multi-block
triangulations using finite element methods is a feasible approach which works 
well in practice. Future work might include adding higher order and/or spectral
elements to this approach.

%=============================================================================
\section*{Acknowledgements}
% -----------------------------------------------------------------------------
This research was supported in part by NSF grant PHY 0505761 to
Louisiana State University and the Teragrid allocation TG-MCA02N014.
The research employed the resources of the CCT at LSU, which is supported by
funding from the Louisiana Legislature's Information Technology
Initiative. 
M. Holst was supported in part by NSF Awards~0715146 and 0511766, 
and DOE Awards DE-FG02-05ER25707 and DE-FG02-04ER25620

We thank Erik Schnetter for helpful discussions throughout this project. 
MT thanks Saul Teukolsky for hospitality at Cornell University,
where part of this work was done. 

We used the Cactus Computational Toolkit \cite{Goodale02a,cactusweb1} with a 
number of locally developed  thorns, the Carpet infrastructure
\cite{Schnetter-etal-03b,carpetweb}, the LAPACK \cite{laug,lapackweb} and 
BLAS \cite{blasweb} libraries from the Netlib Repository \cite{netlibweb}, and the LAM
  \cite{burns94:_lam,squyres03:_compon_archit_lam_mpi,lamweb} and 
MPICH \cite{Gropp:1996:HPI,mpich-user,mpichweb} MPI \cite{mpiweb} implementations.

\bibliographystyle{elsart-num}
\bibliography{fetkbw,references}

\end{document}